\documentclass[apj,twocolappendix]{emulateapj}

\usepackage{times}
\usepackage{balance}

\shorttitle{Transition Region Abundances}
\shortauthors{Warren, Brooks, Doschek, \& Feldman}

\begin{document}


\title{Transition Region Abundance Measurements During Impulsive Heating Events}

\author{Harry P. Warren\altaffilmark{1}, David H. Brooks\altaffilmark{2,3}, George
  A. Doschek\altaffilmark{1}, Uri Feldman\altaffilmark{4}}

\affiliation{\altaffilmark{1}Space Science Division, Naval Research Laboratory, Washington, DC
  20375 USA}

\affiliation{\altaffilmark{2}College of Science, George Mason University, 4400 University Drive,
  Fairfax, VA 22030 USA}

\affiliation{\altaffilmark{4}Artep, Inc., 2922 Excelsior Springs Court, Ellicott City, MD 21042,
  USA}

\altaffiltext{3}{Present address: Hinode Team, ISAS/JAXA, 3-1-1 Yoshinodai, Chuo-ku, Sagamihara,
  Kanagawa 252-5210, Japan}


\begin{abstract}
 It is well established that elemental abundances vary in the solar atmosphere and that this
 variation is organized by first ionization potential (FIP). Previous studies have shown that in
 the solar corona low-FIP elements, such as Fe, Si, Mg, and Ca, are generally enriched relative to
 high-FIP elements, such as C, N, O, Ar, and Ne. In this paper we report on measurements of plasma
 composition made during impulsive heating events observed at transition region temperatures with
 the Extreme Ultraviolet Imaging Spectrometer (EIS) on \textit{Hinode}. During these events the
 intensities of O\,\textsc{iv}, \textsc{v}, and \textsc{vi} emission lines are enhanced relative
 to emission lines from Mg\,\textsc{v}, \textsc{vi}, and \textsc{vii} and Si\,\textsc{vi} and
 \textsc{vii} and indicate a composition close to that of the photosphere.  Long-lived coronal fan
 structures, in contrast, show an enrichment of low-FIP elements. We conjecture that the plasma
 composition is an important signature of the coronal heating process, with impulsive heating
 leading to the evaporation of unfractionated material from the lower layers of the solar
 atmosphere and higher frequency heating leading to long-lived structures and the accumulation of
 low-FIP elements in the corona.
\end{abstract}



\section{introduction}

Understanding how mass and energy flows through the solar upper atmosphere is one of the most
complex problems in astrophysics.  One important clue to understanding this process lies in the
fact that the plasma composition is not constant throughout the solar atmosphere, but varies from
feature to feature. These variations are organized by first ionization potential (FIP), elements
with a FIP below 10\,eV, such as Fe, Si, Mg, and Ca, are generally enriched in the solar corona,
while elements with higher FIP, such as C, N, O, Ar, and Ne, have the composition of the
photosphere (for a review of measurements and models see \citealt{laming2015}). The fact that this
behavior depends on the first ionization potential suggests that the fractionation occurs in the
solar chromosphere, where the high-FIP elements are neutral and the low-FIP elements are ionized.

Observations of high temperature, active region plasma indicate an enrichment of 3--4 for low-FIP
elements. For example, \citet{delzanna2014}, reanalyzed soft X-ray observations from the Solar
Maximum Mission (SMM) and found an enrichment of about 3.2 in non-flaring active
regions. \citet{delzanna2013} derived a similar value for the enrichment of low-FIP elements in
active region emission using observations from the Extreme Ultraviolet Imaging Spectrometer (EIS)
on \textit{Hinode}. \citet{warren2012} used EIS observations to measure the temperature structure
of 15 active regions and found the relative intensities of high-FIP lines to be consistent with a
coronal composition. Furthermore, these analyses found relatively narrow temperature distributions
in the core of the active region near 3--4\,MK.

Recent measurements in solar flares have generally indicated abundances that are much closer to
that of the photosphere. \citet{dennis2015}, \cite{fludra1999}, and \citet{warren2014a}, for
example, found enrichments of $1.66\pm0.34$, $1.4\pm0.4$, $1.17\pm0.22$, respectively, of high
temperature Fe lines observed in a large sample of solar flares. \citet{sylwester2015} derived
abundances close to photospheric for high temperature Si, Ar, and S lines observed during
flares. In many of these studies the temperature distributions are observed to be very broad with
a peak near 10\,MK.

The differences in composition and temperature distribution between high temperature active region
plasma and high temperature solar flare plasma suggest that they may be heated in different
ways. One possible scenario is that active region loops are heated by frequent, small scale events
that keep these loops relatively close to equilibrium. Individual flare loops, in contrast, could
be heated impulsively and are always evolving. This would account for the differences in
temperature distributions (narrow vs.\ broad) and the composition (coronal vs.\ photospheric),
since it appears that the fractionation process takes several days to enrich the corona with low-FIP
elements \citep[e.g.,][]{sheeley1995,widing2001}.

Unfortunately, not all of the observational evidence is consistent with this simple
picture. \citet{dennis2015} and \citet{doschek2015}, for example, find that Ca is enriched at all
times during flares. \citet{sylwester2015} find evidence for a similar enrichment in K
emission. \citet{doschek2015} have even identified isolated regions in several flares where Ar
emission is strongly enhanced relative to Ca. 

It is clear that additional observations are needed to make progress on understanding the
relationship between plasma composition and the coronal heating process. It has long been
recognized that the transition region, the interface between the hot corona and the mass reservoir
of the chromosphere, should provide important information. The complexity of the solar transition
region emission and the high cadence, high spatial resolution, and broad wavelength coverage
needed to observe it, however, have largely thwarted this approach.

In this paper we present the analysis of transition region emission observed with the EIS
spectrometer on \textit{Hinode}. At the EUV wavelengths observed by EIS, the transition region
emission lines are weak. They can, however, be observed easily during intense heating events such
as flares. We find that during these events the observed intensities are generally consistent with
a plasma composition close to that of the photosphere. For long-lived coronal ``fans, '' in
contrast, the line intensities are more consistent with a coronal composition. 
\section{observations}

\begin{deluxetable}{rrrrrrr}
\tabletypesize{\scriptsize}
\tablecaption{EIS Emission Lines of Interest\tablenotemark{a}}
\tablehead{
  \multicolumn{1}{c}{Line} &
  \multicolumn{1}{c}{$\log T_{\ast}$} &
  \multicolumn{1}{c}{$\log n_{\ast}$} &
  \multicolumn{1}{c}{$\epsilon(T_\ast, n_\ast)$} &
  \multicolumn{1}{c}{FIP} &
  \multicolumn{1}{c}{$A_p$} &
  \multicolumn{1}{c}{$A_c$}
}  
\startdata
     \ion{O}{4} 279.933 &     5.26 &    10.44 &   1.45e-25 &     13.6 &     8.76 &     8.76 \\
     \ion{O}{5} 248.456 &     5.42 &    10.28 &   1.90e-25 &     13.6 &     8.76 &     8.76 \\
    \ion{Mg}{5} 276.579 &     5.50 &    10.20 &   5.13e-26 &      7.6 &     7.54 &     8.14 \\
     \ion{O}{6} 184.117 &     5.52 &    10.18 &   5.90e-26 &     13.6 &     8.76 &     8.76 \\
    \ion{Si}{6} 246.004 &     5.64 &    10.06 &   5.44e-26 &      8.2 &     7.52 &     8.12 \\
    \ion{Mg}{6} 268.986 &     5.66 &    10.04 &   2.32e-26 &      7.6 &     7.54 &     8.14 \\
    \ion{Mg}{7} 276.153 &     5.80 &     9.90 &   1.15e-26 &      7.6 &     7.54 &     8.14 \\
    \ion{Mg}{7} 280.737 &     5.78 &     9.92 &   4.02e-26 &      7.6 &     7.54 &     8.14 \\
    \ion{Si}{7} 275.368 &     5.80 &     9.90 &   8.06e-26 &      8.2 &     7.52 &     8.12 \\
    \ion{Fe}{9} 197.862 &     5.96 &     9.74 &   1.98e-26 &      7.9 &     7.52 &     8.12 \\
   \ion{Fe}{10} 184.536 &     6.06 &     9.64 &   1.06e-25 &      7.9 &     7.52 &     8.12 \\
   \ion{Fe}{11} 180.401 &     6.14 &     9.56 &   3.29e-25 &      7.9 &     7.52 &     8.12 \\
   \ion{Si}{10} 258.375 &     6.14 &     9.56 &   9.92e-26 &      8.2 &     7.52 &     8.12 \\
    \ion{S}{10} 264.233 &     6.18 &     9.52 &   3.33e-26 &     10.4 &     7.16 &     7.16 \\   
   \ion{Fe}{12} 192.394 &     6.20 &     9.50 &   8.08e-26 &      7.9 &     7.52 &     8.12 \\
   \ion{Fe}{13} 202.044 &     6.26 &     9.44 &   9.16e-26 &      7.9 &     7.52 &     8.12 \\
   \ion{Fe}{13} 203.826 &     6.24 &     9.46 &   2.09e-25 &      7.9 &     7.52 &     8.12 \\
   \ion{Fe}{14} 274.203 &     6.30 &     9.40 &   9.75e-26 &      7.9 &     7.52 &     8.12 \\
   \ion{Fe}{15} 284.160 &     6.34 &     9.36 &   5.55e-25 &      7.9 &     7.52 &     8.12 \\
   \ion{Fe}{16} 262.984 &     6.42 &     9.28 &   2.19e-26 &      7.9 &     7.52 &     8.12
   \enddata
\tablenotetext{a}{Plasma emissivity as a function of temperature is shown in
  Figure~\ref{fig:atodat}. Here we show the peak temperature ($T_{\ast}$), the number density at
  the peak temperature ($n_{\ast}$), and the emissivity computed for these parameters assuming a
  photospheric composition. Also shown for each ion are the FIP in eV and the logarithmic
  abundances relative to H ($A_H=12$). The photospheric abundances ($A_p$) are those tabulated by
  \citet{caffau2011}. For coronal abundances ($A_c$) we approximate the tabulation of
  \citet{feldman1992} by scaling the photospheric abundances of the low FIP elements by a factor
  of 4 and leaving the abundances of the high FIP elements unchanged. }
\label{table:atodat}   
\end{deluxetable}

EIS \citep{culhane2007,korendyke2006} is a high spatial and spectral resolution imaging
spectrograph. EIS observes two wavelength ranges, 171--212\,\AA\ and 245--291\,\AA, with a
spectral resolution of about 22\,m\AA\ and a spatial resolution of about 1\arcsec\ per pixel.
Solar images can be made by stepping the slit over a region of the Sun and taking an exposure at
each position.

\begin{figure}[t!]
\centerline{\includegraphics[clip,angle=90,width=\linewidth]{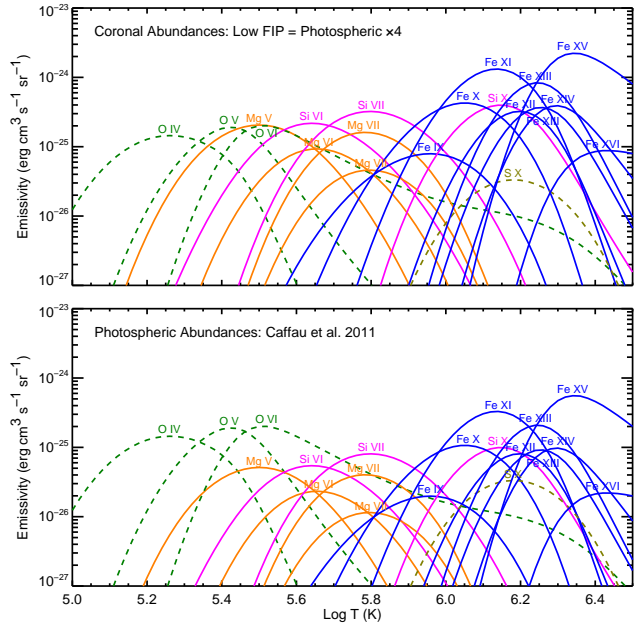}}
 \caption{Plasma emissivity as a function of temperature for the EIS emission lines of interest.
   A constant pressure of $10^{16}$\,cm$^{-3}$~K has been assumed.  Calculations are shown for
   photospheric and coronal abundances. Emissivities for high-FIP elements are shown with dashed
   lined. These calculations suggest that \ion{O}{5} and \ion{O}{6} emission lines should be
   bright relative to \ion{Mg}{5}, \ion{Mg}{6}, and \ion{Si}{6} when the composition is
   photospheric.  }
\label{fig:atodat}
\end{figure}

Some information on the EIS emission lines of interest for this study is shown in
Table~\ref{table:atodat} and in Figure~\ref{fig:atodat}. These data are taken from the CHIANTI
atomic database version 8.0.1 \citep{delzanna2015b,dere1997}. As can be seen in the plots of
plasma emissivity in Figure~\ref{fig:atodat}, none of the transition region emission lines that
can be observed with EIS are so well matched in temperature that we can use a simple line ratio to
measure the composition, as was done with the \ion{Mg}{6} and \ion{Ne}{6} lines observed with
Skylab \citep[e.g.,][]{sheeley1995,widing2001}. Instead, we will use a series of O, Mg, and Si
emission lines to infer the general trends in the relative intensities between low and high-FIP
elements.

In this paper we use the tabulation of \citet{caffau2011} for the elemental abundances of the
photosphere. For coronal abundances we approximate the tabulation of \citet{feldman1992} by
scaling the photospheric abundances of the low-FIP elements by a factor of 4 and leaving the
abundances of the high-FIP elements unchanged. Adopting the abundances of \citet{feldman1992}
would imply changes in both the high and low-FIP elements in the corona, although the difference
for the high-FIP elements would be relatively small. The relevant abundances are given in
Table~\ref{table:atodat}.

For the emissivities shown in Table~\ref{table:atodat} and in Figure~\ref{fig:atodat} we have
assumed a constant total pressure of $P = 2n_eT_e = 10^{16}$\,cm$^{-3}$~K. To measure electron
densities for our observations we use the \ion{Mg}{7} 280.737/276.153 \citep{landi2009} and
\ion{Fe}{13} 203.826/202.044 line ratios \citep{warren2010}. Note that almost all of the emission
lines of interest have some amount of density sensitivity so estimating the density in the
emitting plasma is an important component of the analysis.

Most of the emission lines selected for this study have been analyzed extensively and are known to
yield consistent results \citep[e.g.,][]{landi2009,brooks2009,warren2009}. Unfortunately, the two
\ion{O}{6} emission lines in the EIS wavelength range, 183.937 and 184.117\,\AA, are
problematic. On the disk the observed 183.937/184.117 ratio differs significantly from theory. The
strong wavelength shifts in the 183.937\,\AA\ line suggest that it is blended
\citep{landi2009}. The observed intensity of the longer wavelength line is not consistent with
other emission lines \citep[e.g.,][]{warren2009}. As can be see in Figure~\ref{fig:atodat},
\ion{O}{6} is needed to improve the overlap between the high and low-FIP transition region
emission lines. For this study we scale the emissivity of \ion{O}{6} 184.117\,\AA\ by a factor of
3.4 in all cases. Additional details on this line and the scaling factor that we have adopted are
discussed in Appendix A.

\begin{figure*}[t!]
  \newlength{\figurewidth}
  \setlength{\figurewidth}{0.95\linewidth}
  \centerline{
   \includegraphics[clip,width=0.63\figurewidth]{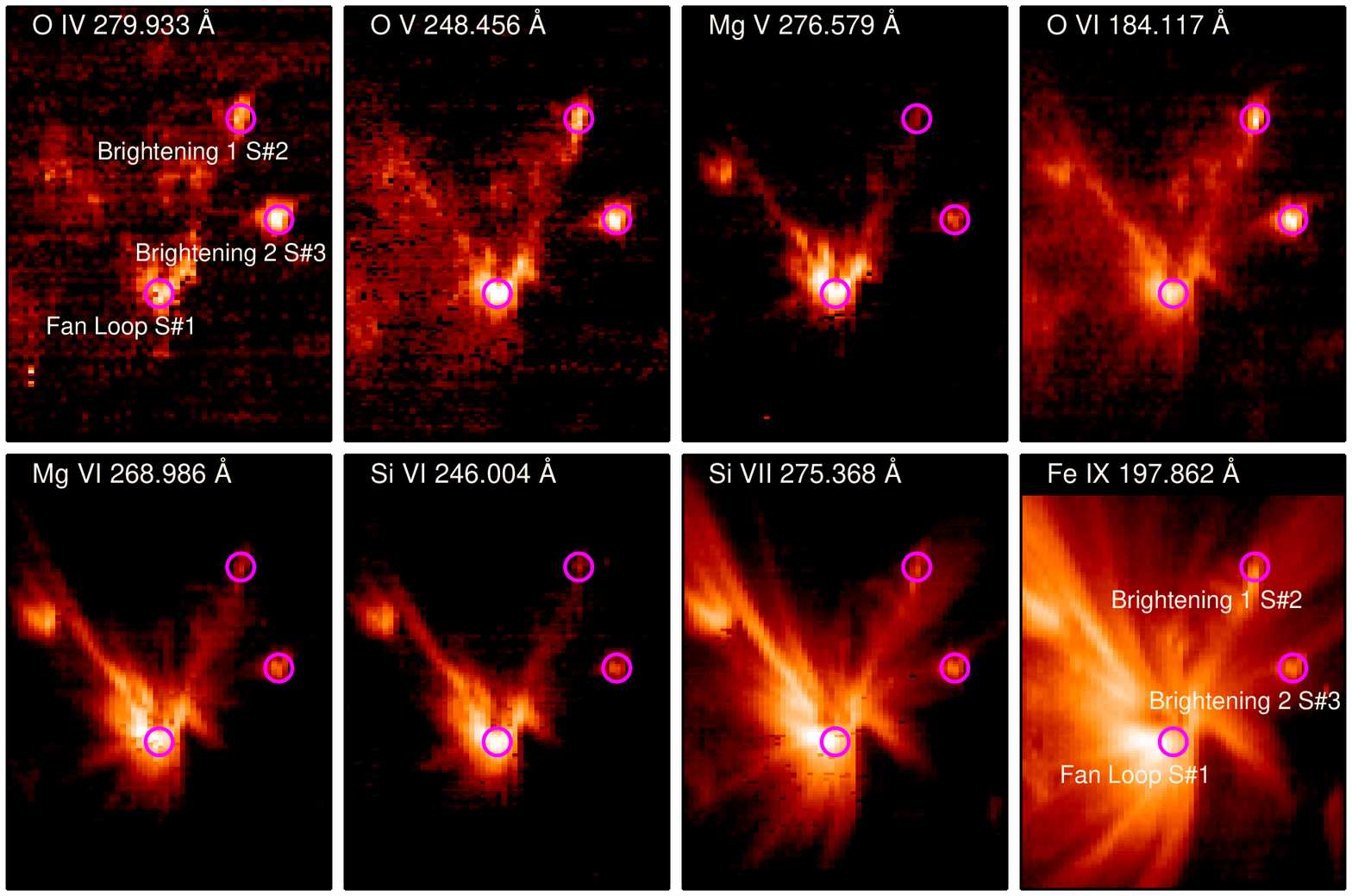}
   \includegraphics[clip,width=0.37\figurewidth]{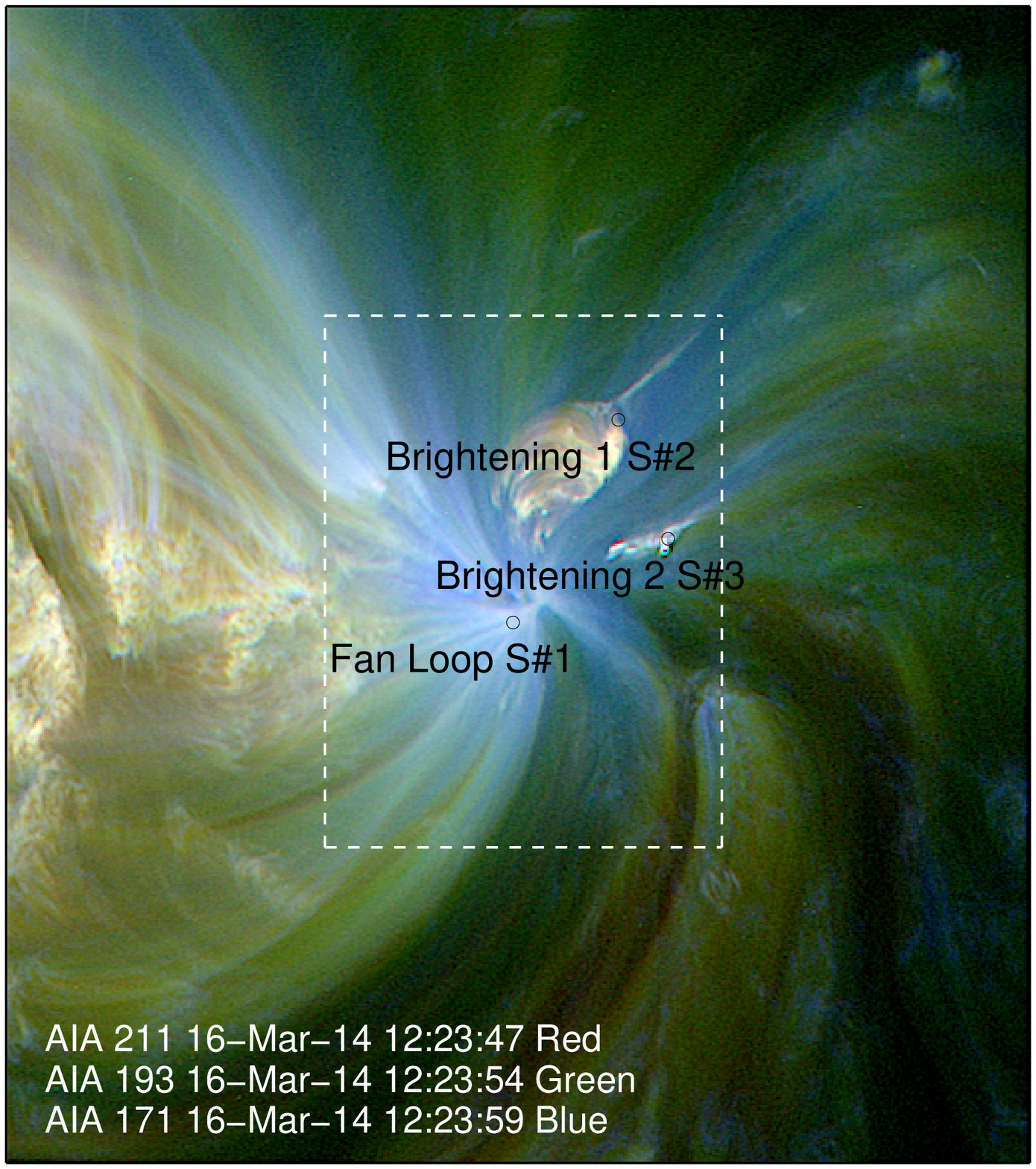}}
  \vspace{-0.1in}
  \centerline{
   \includegraphics[clip,angle=90,width=\figurewidth]{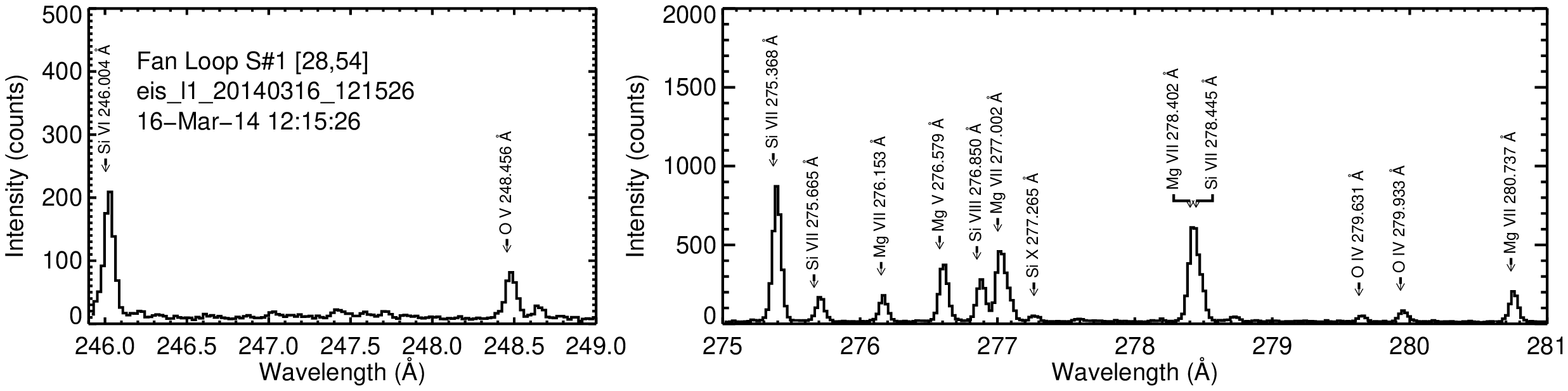}}
  \vspace{-0.2in}  
  \centerline{
   \includegraphics[clip,angle=90,width=\figurewidth]{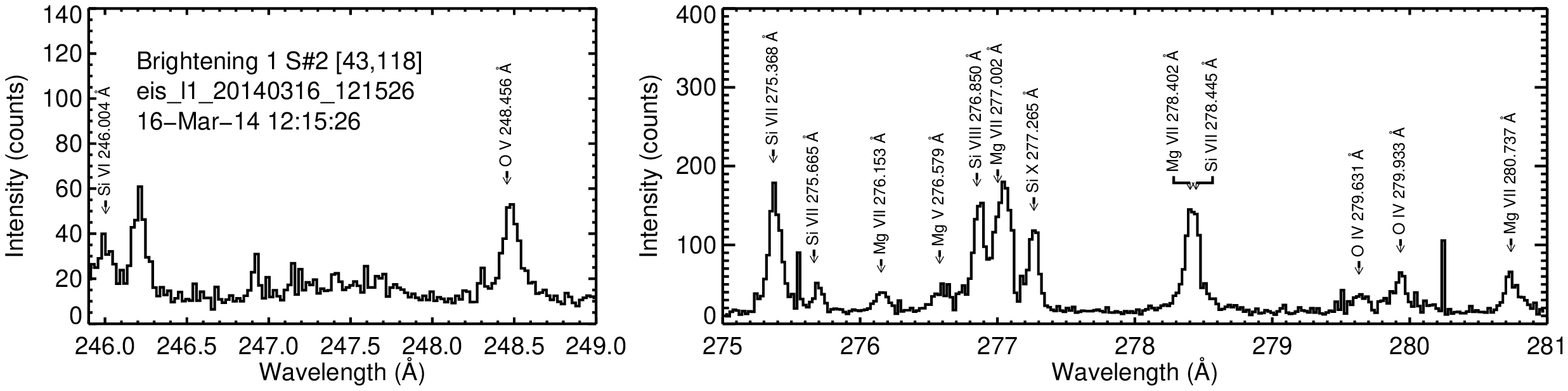}}
  \vspace{-0.2in}    
  \centerline{
   \includegraphics[clip,angle=90,width=\figurewidth]{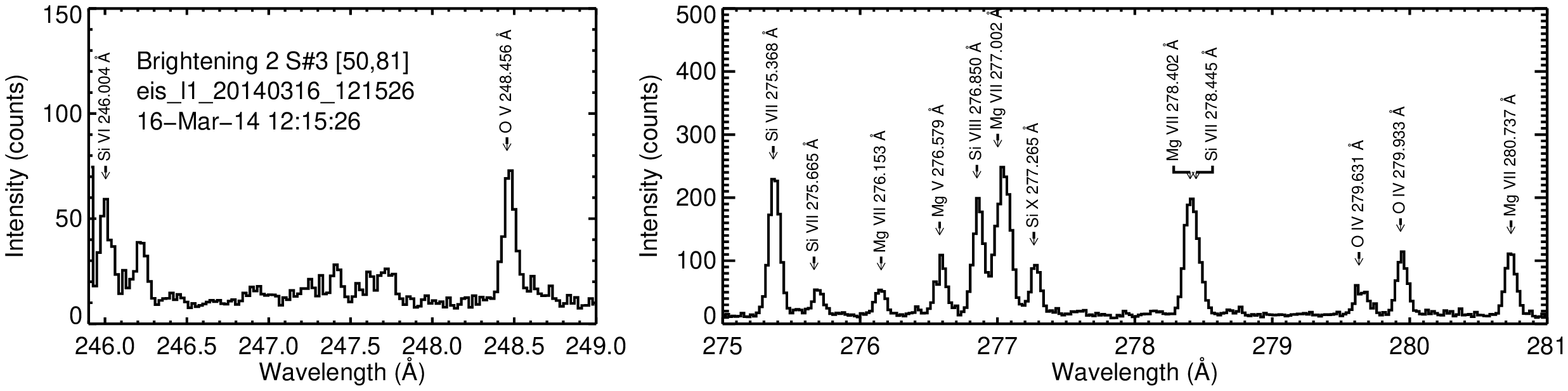}}  
  \caption{EIS and AIA observations of an active region taken on 2014 March 16 between 12:15:26
    and 13:16:05 UT. The top left panels show EIS intensities in various transition region
    emission lines. The top right panel shows a three color AIA image taken during the EIS
    raster. The bottom panels show EIS spectra at selected locations. At the base of the fan loops
    the intensities of the low-FIP Mg and Si lines are bright relative to the high-FIP O lines,
    suggesting a coronal composition. In the transient brightenings the O lines are bright
    relative to Si and Mg, suggesting a photospheric composition. }
\label{fig:r20140316}
\end{figure*}

\begin{figure*}[t!]
  \setlength{\figurewidth}{0.95\linewidth}  
  \centerline{
   \includegraphics[clip,width=0.63\figurewidth]{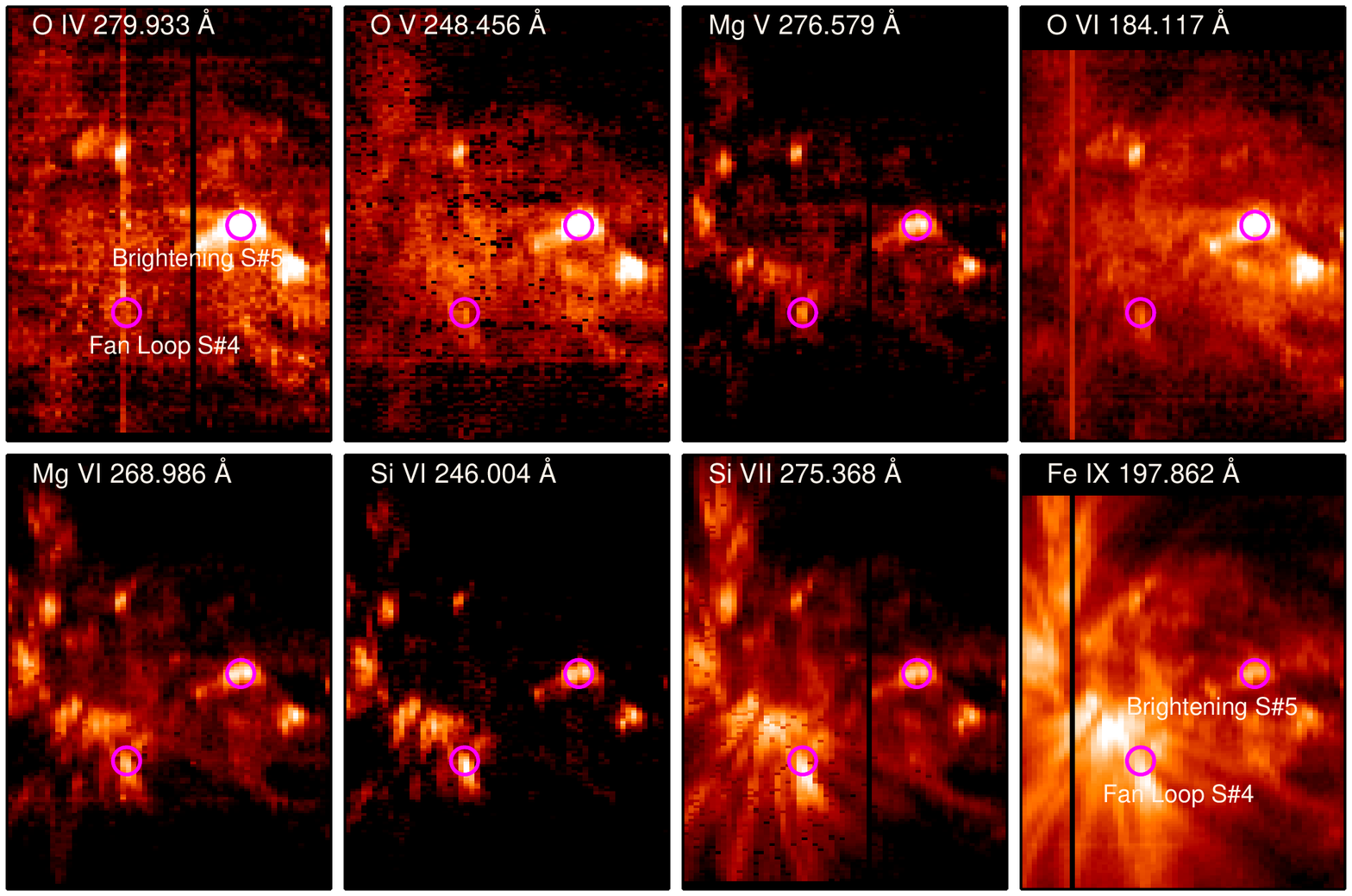}
   \includegraphics[clip,width=0.37\figurewidth]{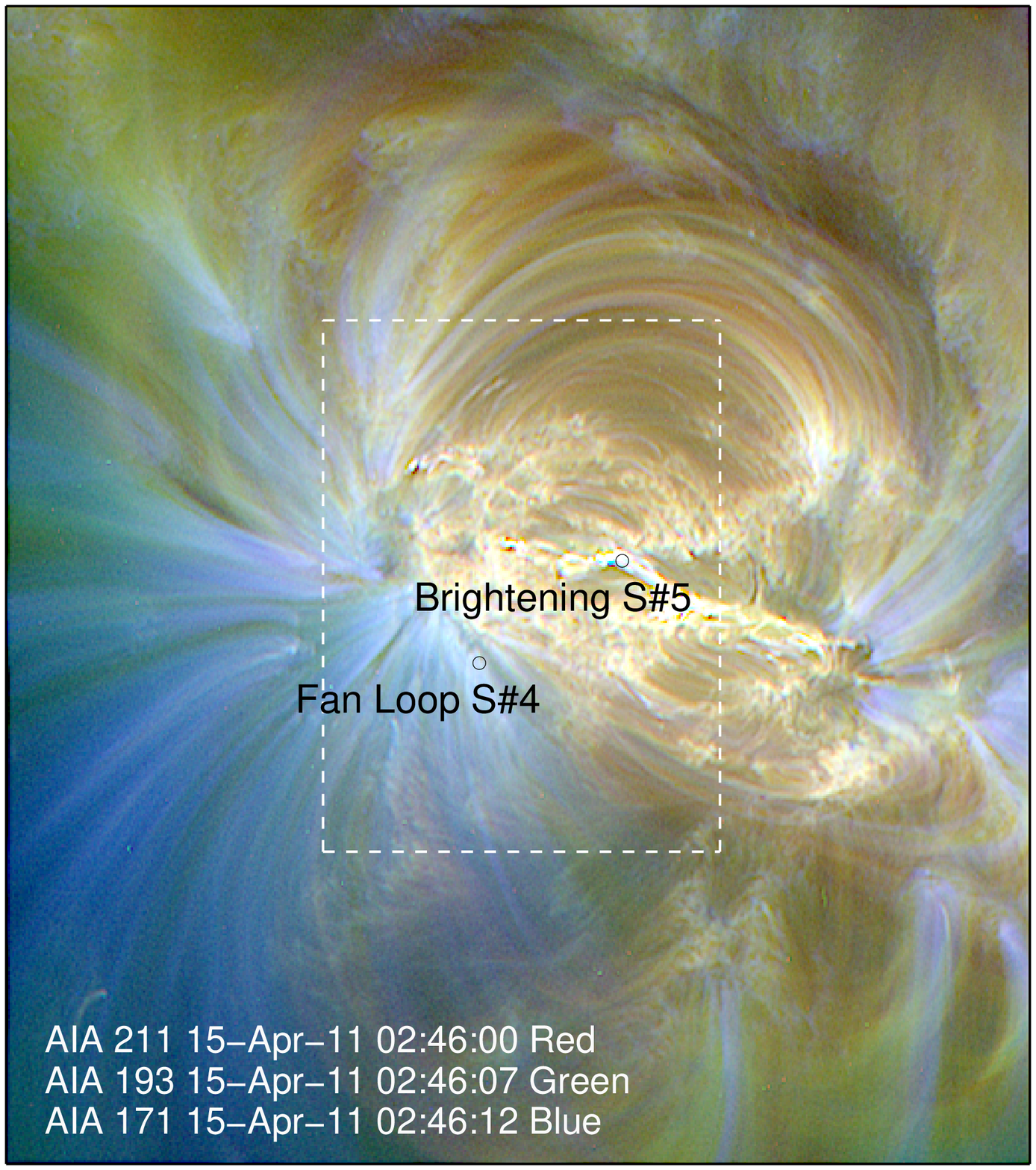}}
  \vspace{-0.1in}
  \centerline{
   \includegraphics[clip,angle=90,width=\figurewidth]{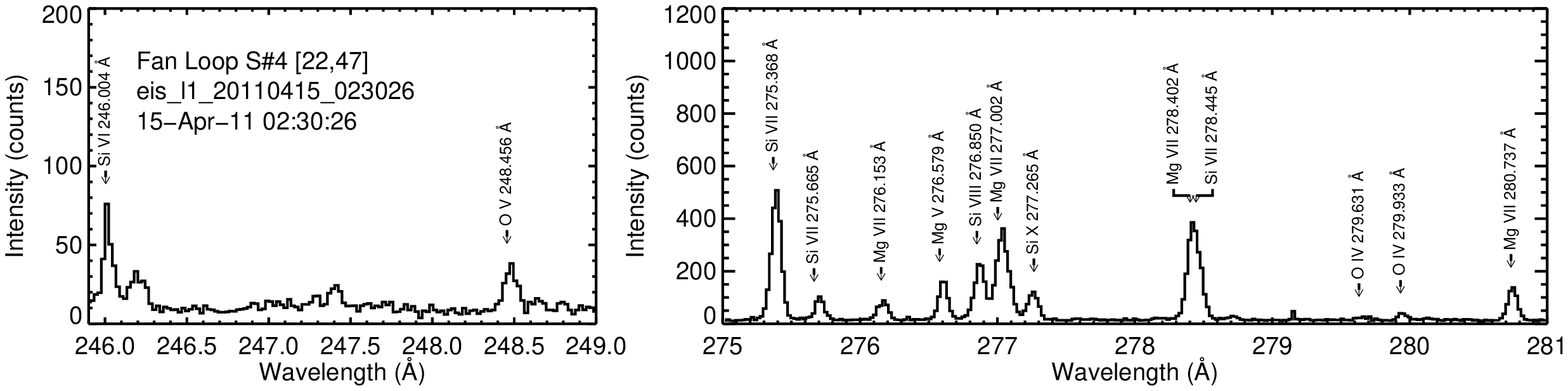}}
  \vspace{-0.2in}  
  \centerline{
   \includegraphics[clip,angle=90,width=\figurewidth]{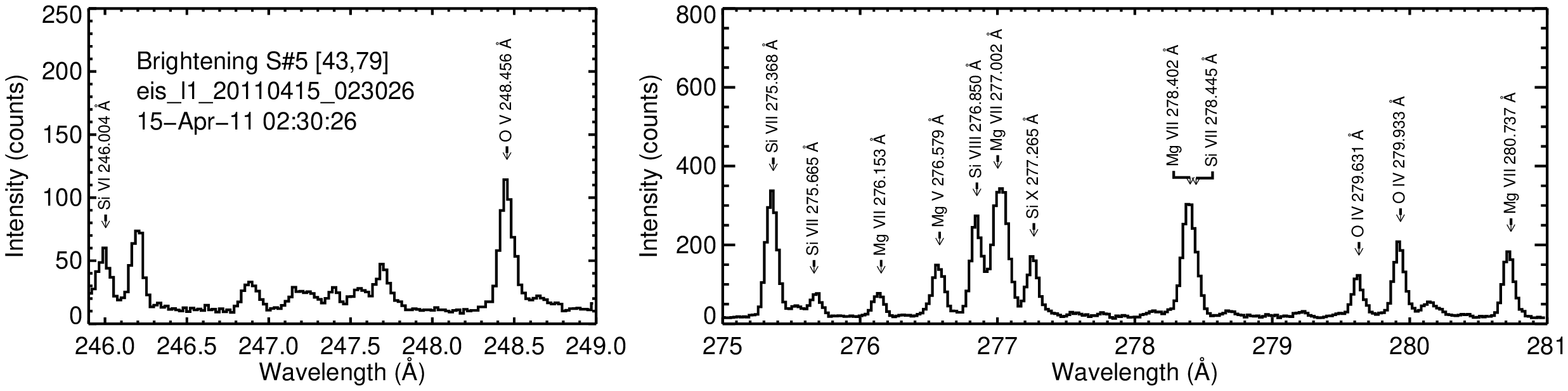}}
  \caption{EIS and AIA observations of an active region taken on 2011 April 15 between 02:30:26
    and 03:31:06 UT. The format is the same as Figure~\ref{fig:r20140316}. In the brightening the
    O emission lines are bright relative to Si and Mg while in the fan loop the relative
    intensities are reversed.}
\label{fig:r20110415}
\end{figure*}

\begin{figure*}[t!]
  \setlength{\figurewidth}{0.95\linewidth}  
  \centerline{
   \includegraphics[clip,width=0.63\figurewidth]{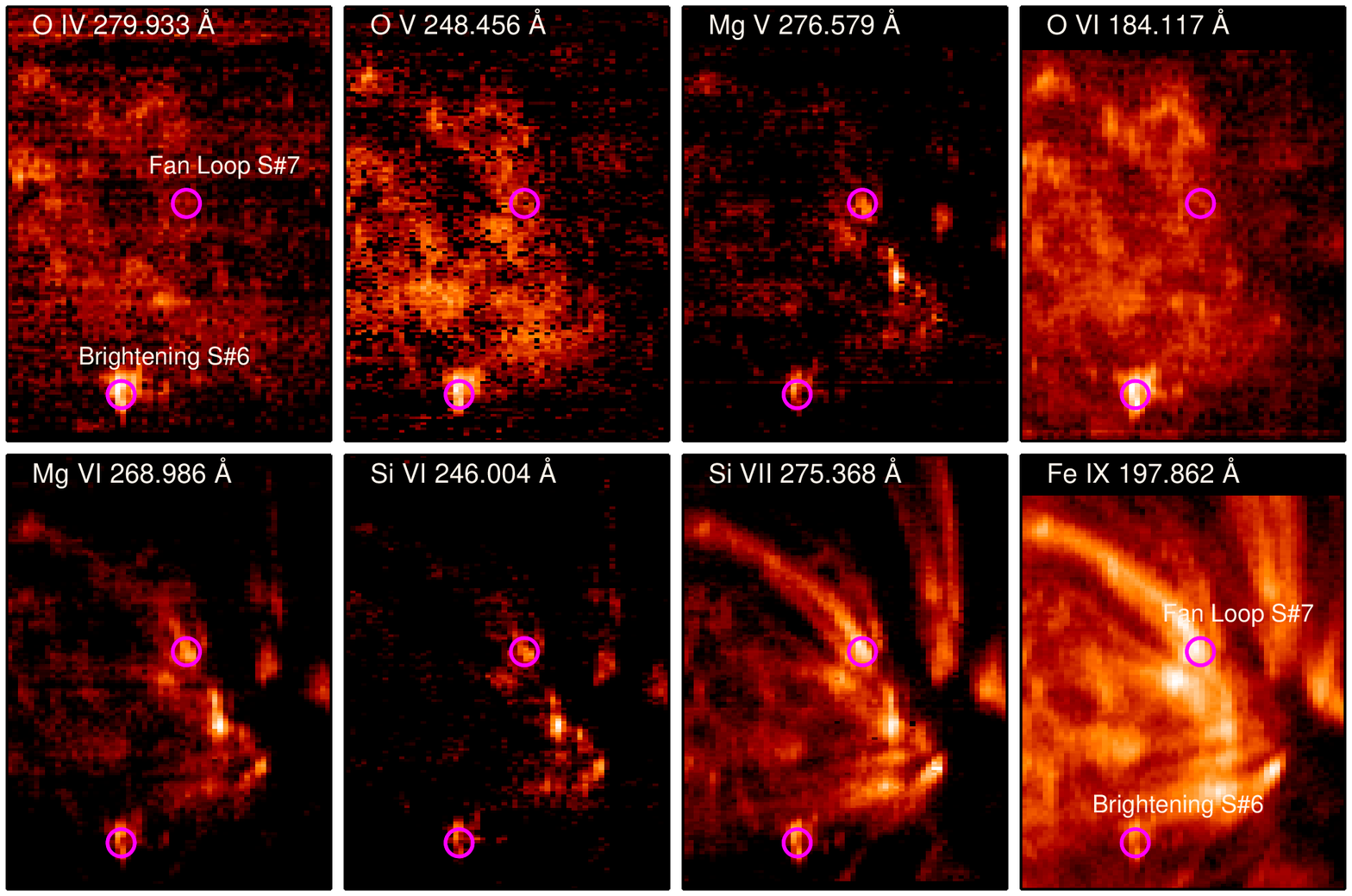}
   \includegraphics[clip,width=0.37\figurewidth]{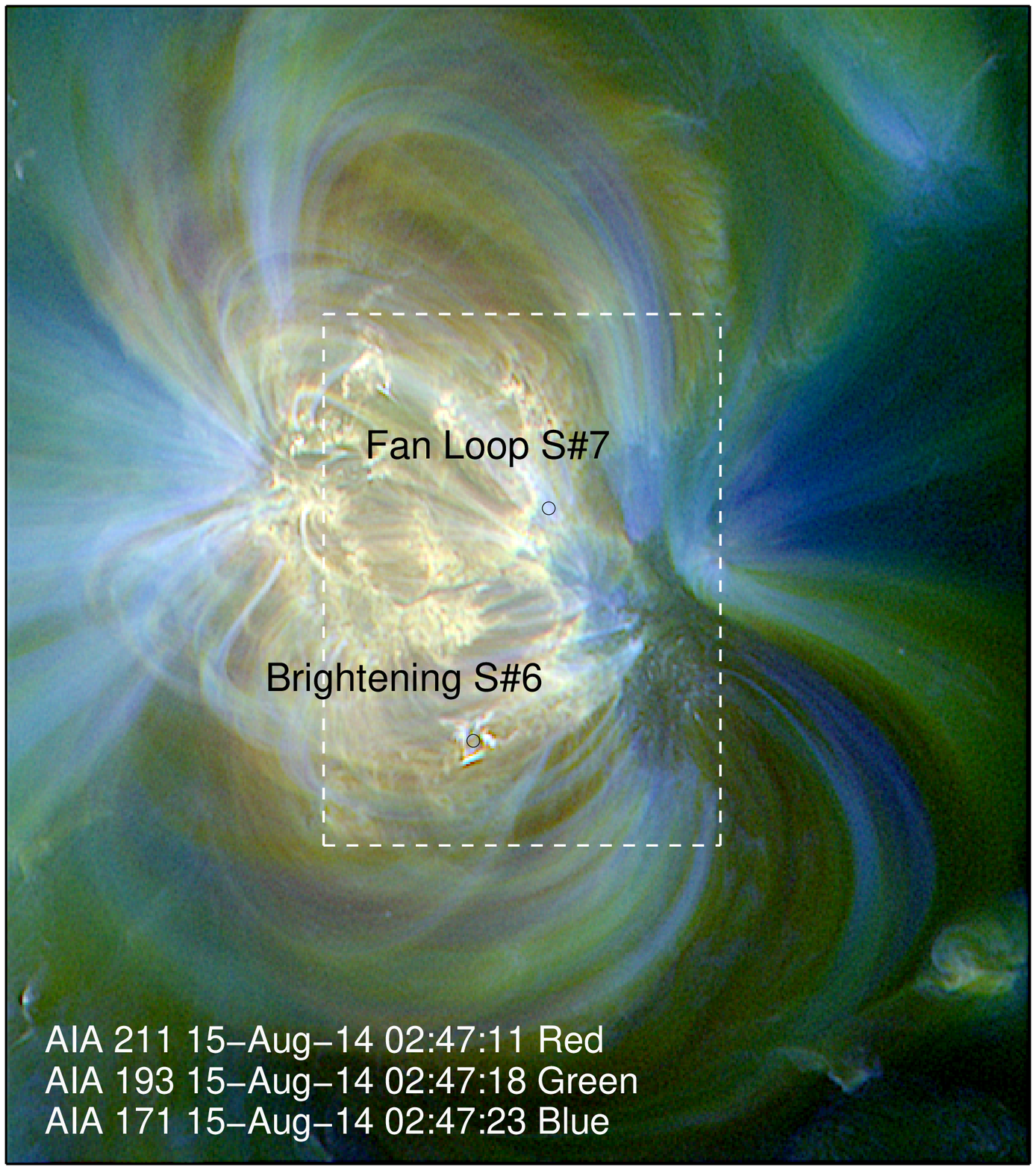}}
  \vspace{-0.1in}
  \centerline{
   \includegraphics[clip,angle=90,width=\figurewidth]{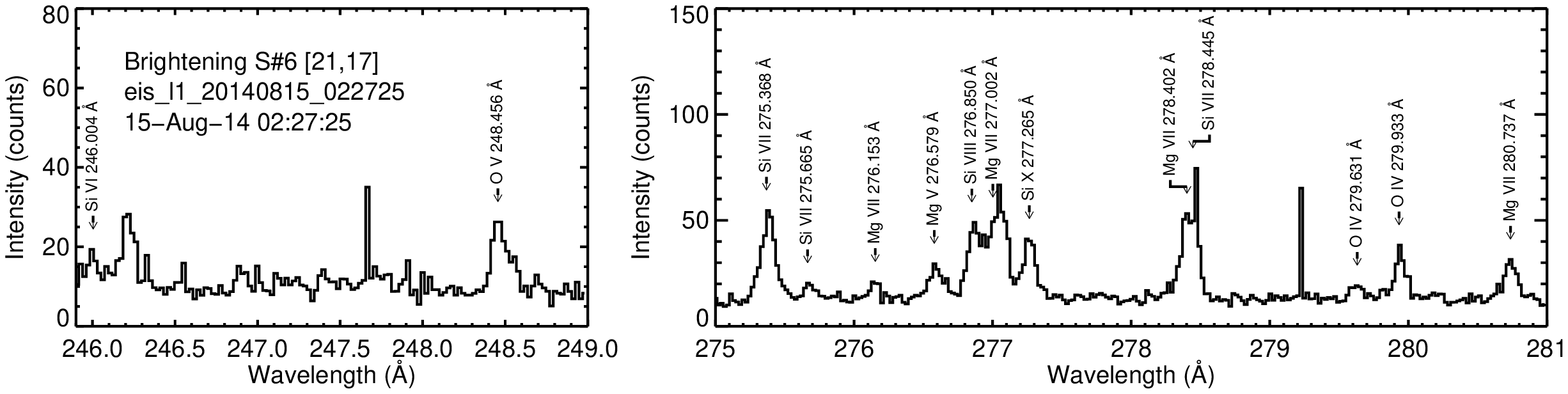}}
  \vspace{-0.2in}  
  \centerline{
   \includegraphics[clip,angle=90,width=\figurewidth]{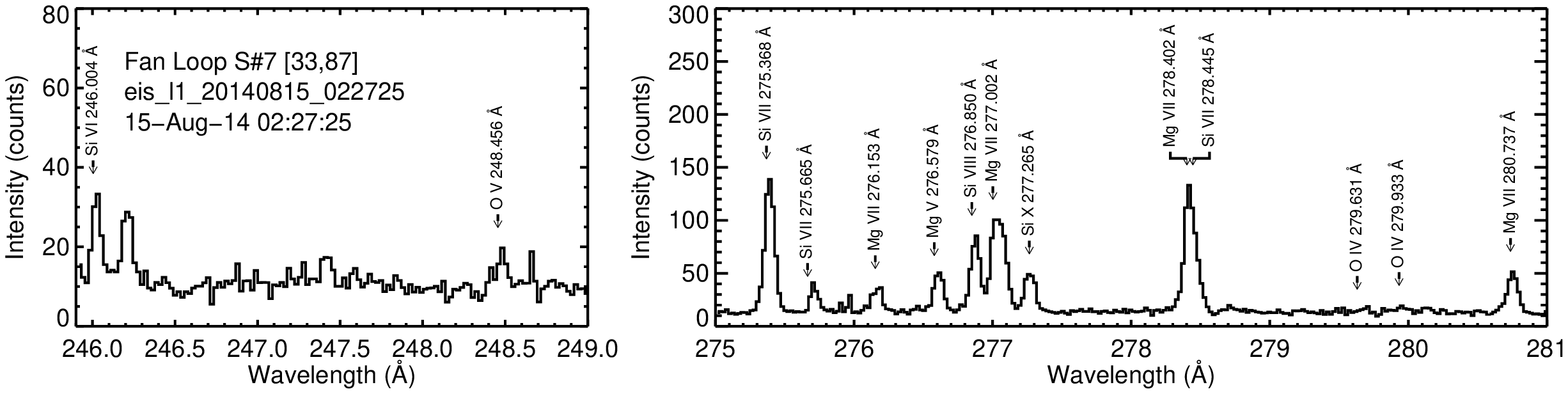}}
  \caption{EIS and AIA observations of an active region taken on 2014 August 15 between 02:27:25
    and 02:58:34 UT. The format is the same as Figure~\ref{fig:r20140316}. Again, in the
    brightening the O emission lines are bright relative to Si and Mg while in the fan loop the
    relative intensities are reversed.}
\label{fig:r20140815}
\end{figure*}

\begin{figure*}[t!]
  \setlength{\figurewidth}{0.95\linewidth}  
  \centerline{
   \includegraphics[clip,width=0.63\figurewidth]{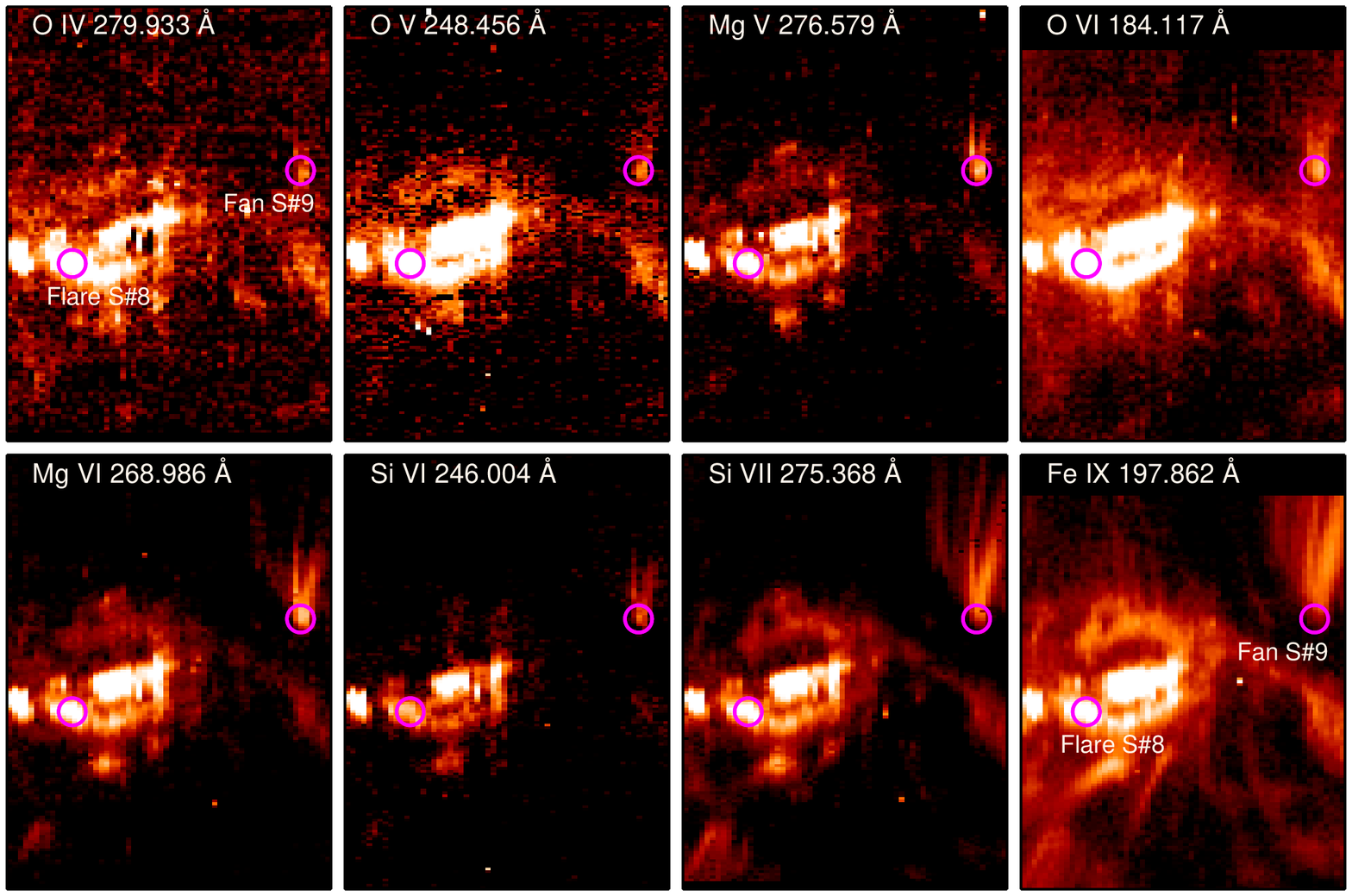}
   \includegraphics[clip,width=0.37\figurewidth]{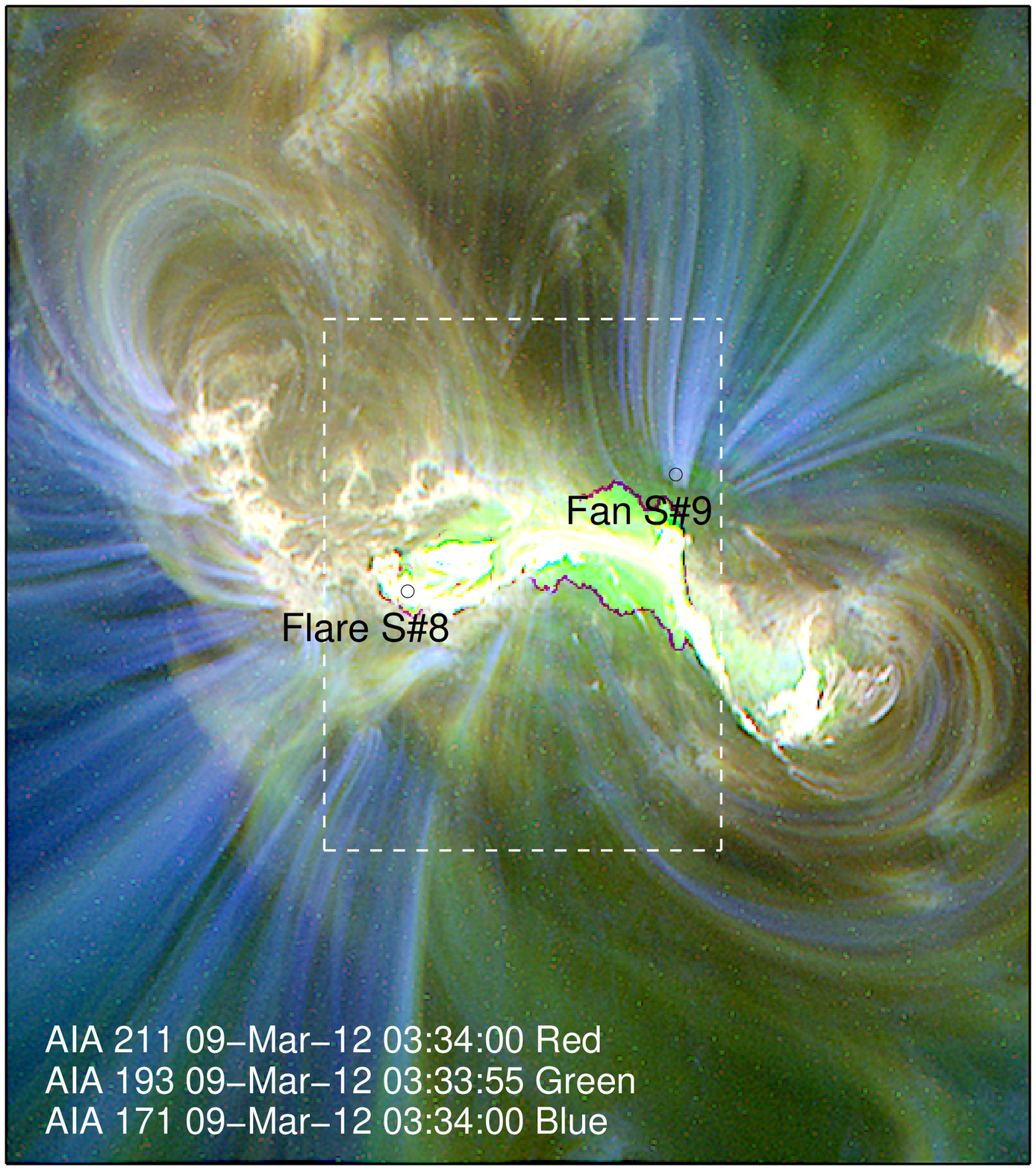}}
  \vspace{-0.1in}
  \centerline{
    \includegraphics[clip,angle=90,width=\figurewidth]{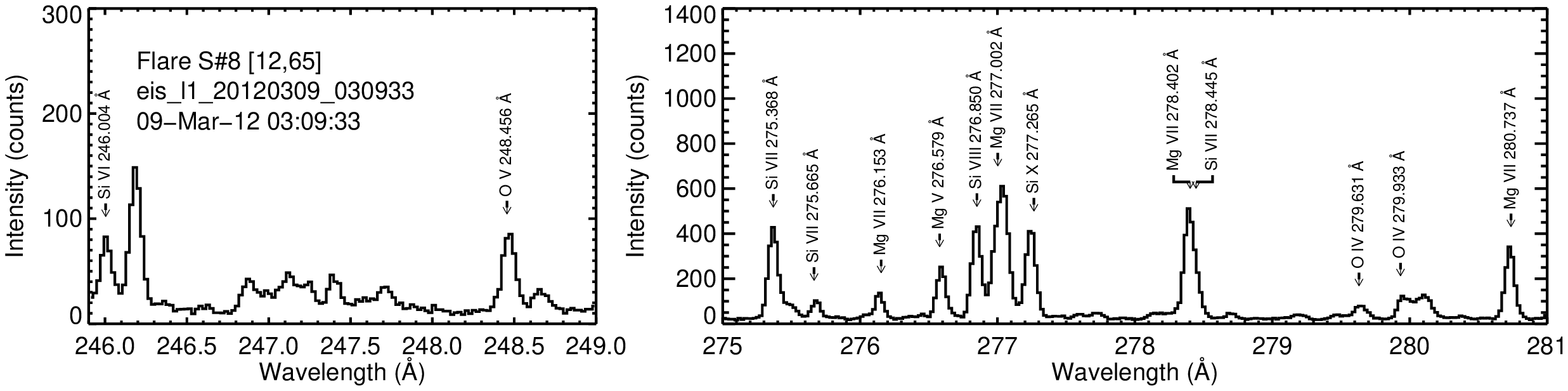}}
  \vspace{-0.2in}  
  \centerline{
   \includegraphics[clip,angle=90,width=\figurewidth]{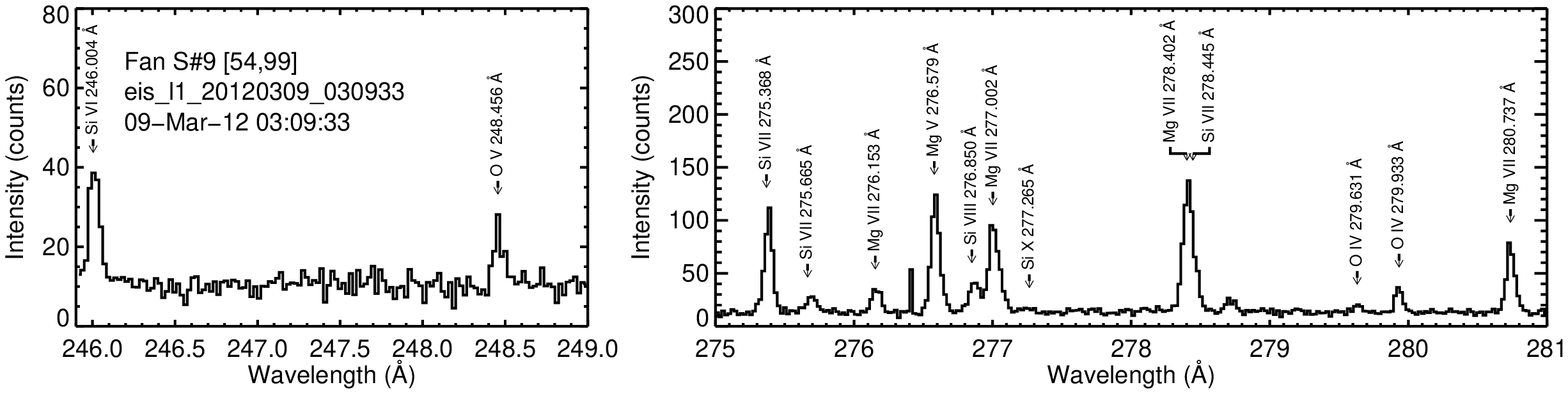}}  
  \caption{EIS and AIA observations of a flaring active region taken on 2012 March 9 between
    03:09:33 and 03:40:42 UT. The format is the same as Figure~\ref{fig:r20140316}.}
\label{fig:r20120309}
\end{figure*}

\begin{figure*}[t!]
  \setlength{\figurewidth}{0.95\linewidth}  
  \centerline{
   \includegraphics[clip,width=0.63\figurewidth]{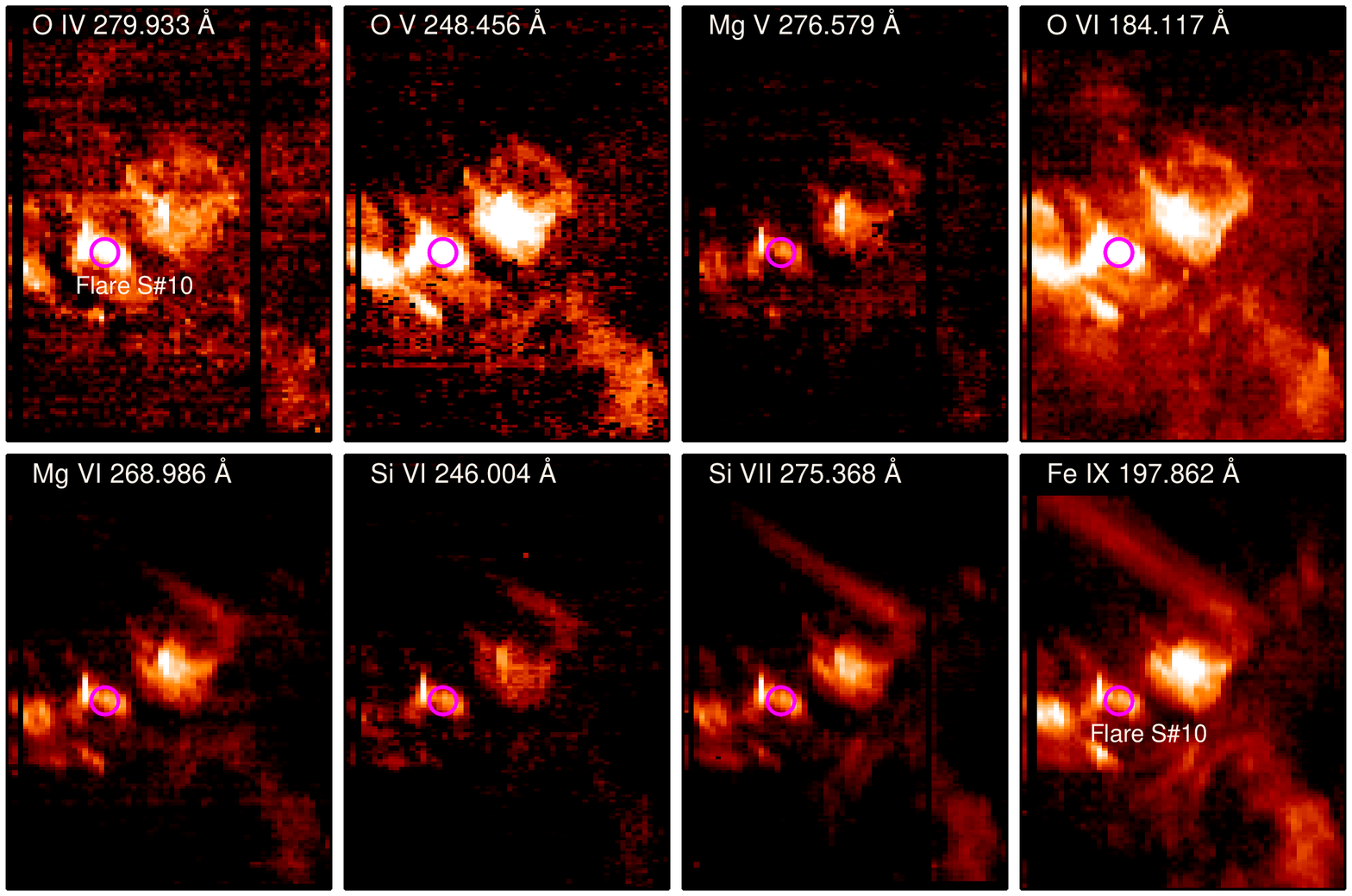}
   \includegraphics[clip,width=0.37\figurewidth]{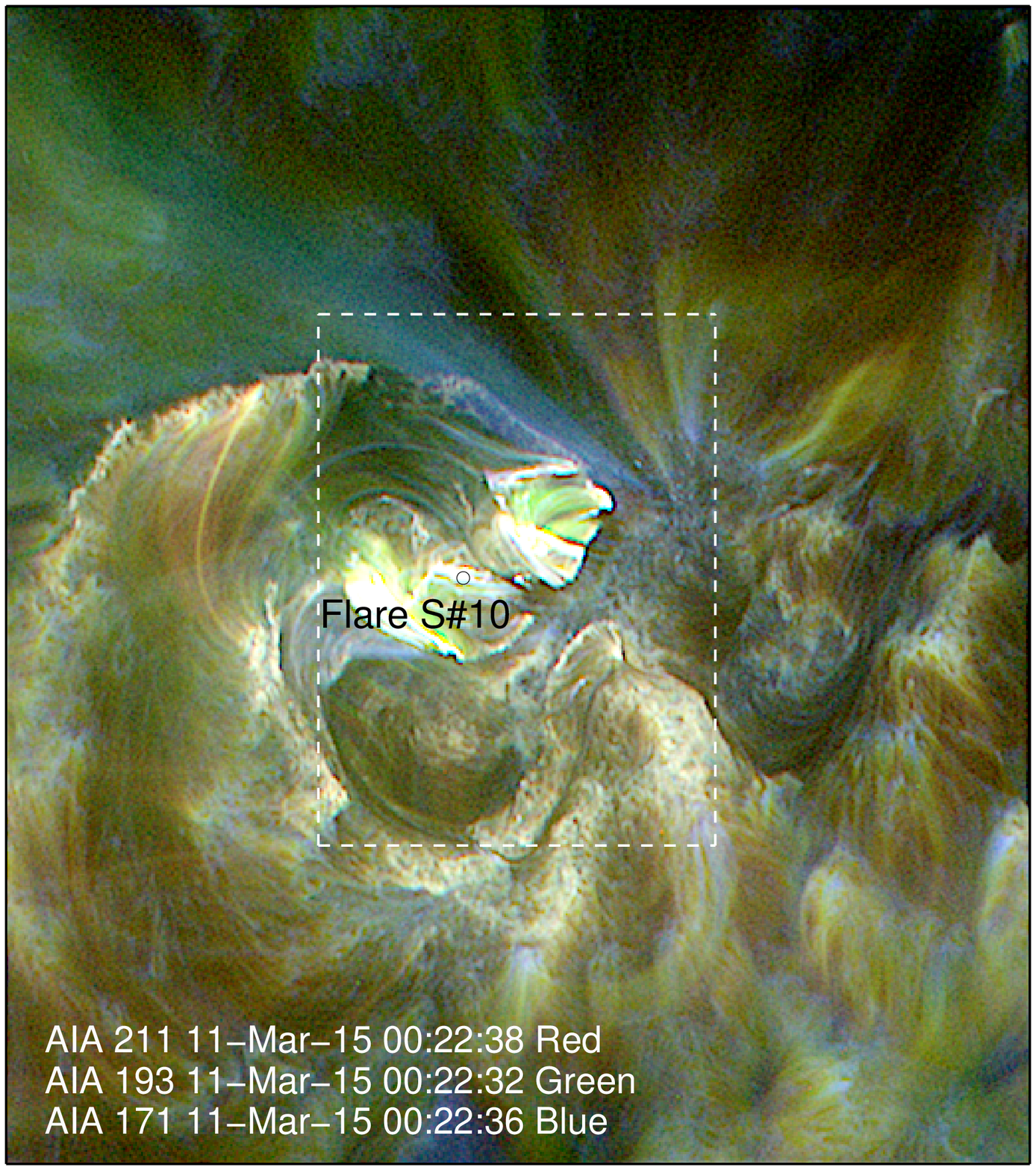}}
  \vspace{-0.1in}
  \centerline{
    \includegraphics[clip,angle=90,width=\figurewidth]{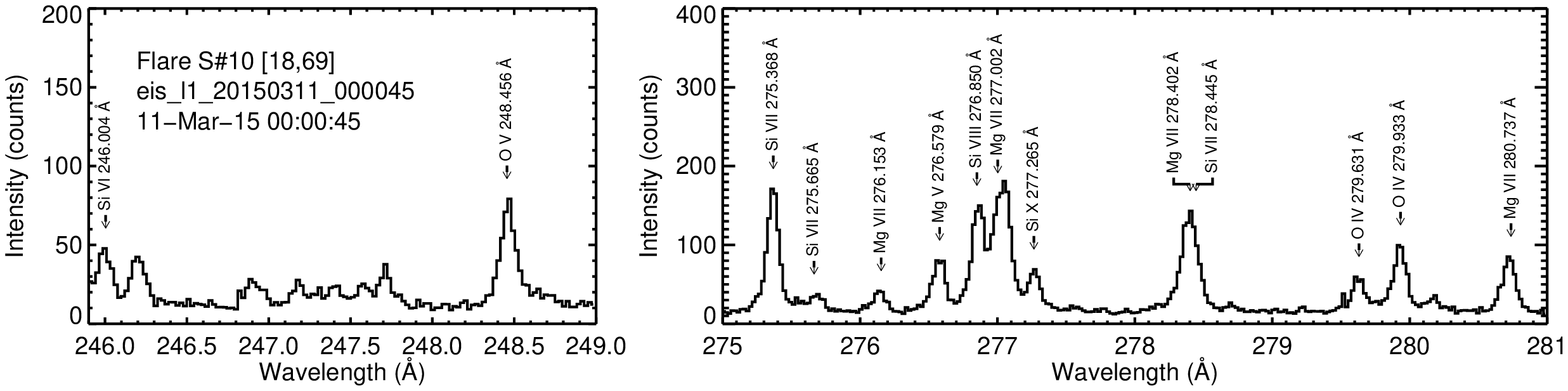}}
  \caption{EIS and AIA observations of a flaring active region taken on 2015 March 11 between
    00:00:45 and 00:31:54 UT. The format is the same as Figure~\ref{fig:r20140316}.}
\label{fig:r20150311}
\end{figure*}

\begin{figure*}[t!]
  \setlength{\figurewidth}{0.95\linewidth}  
  \centerline{
   \includegraphics[clip,width=0.63\figurewidth]{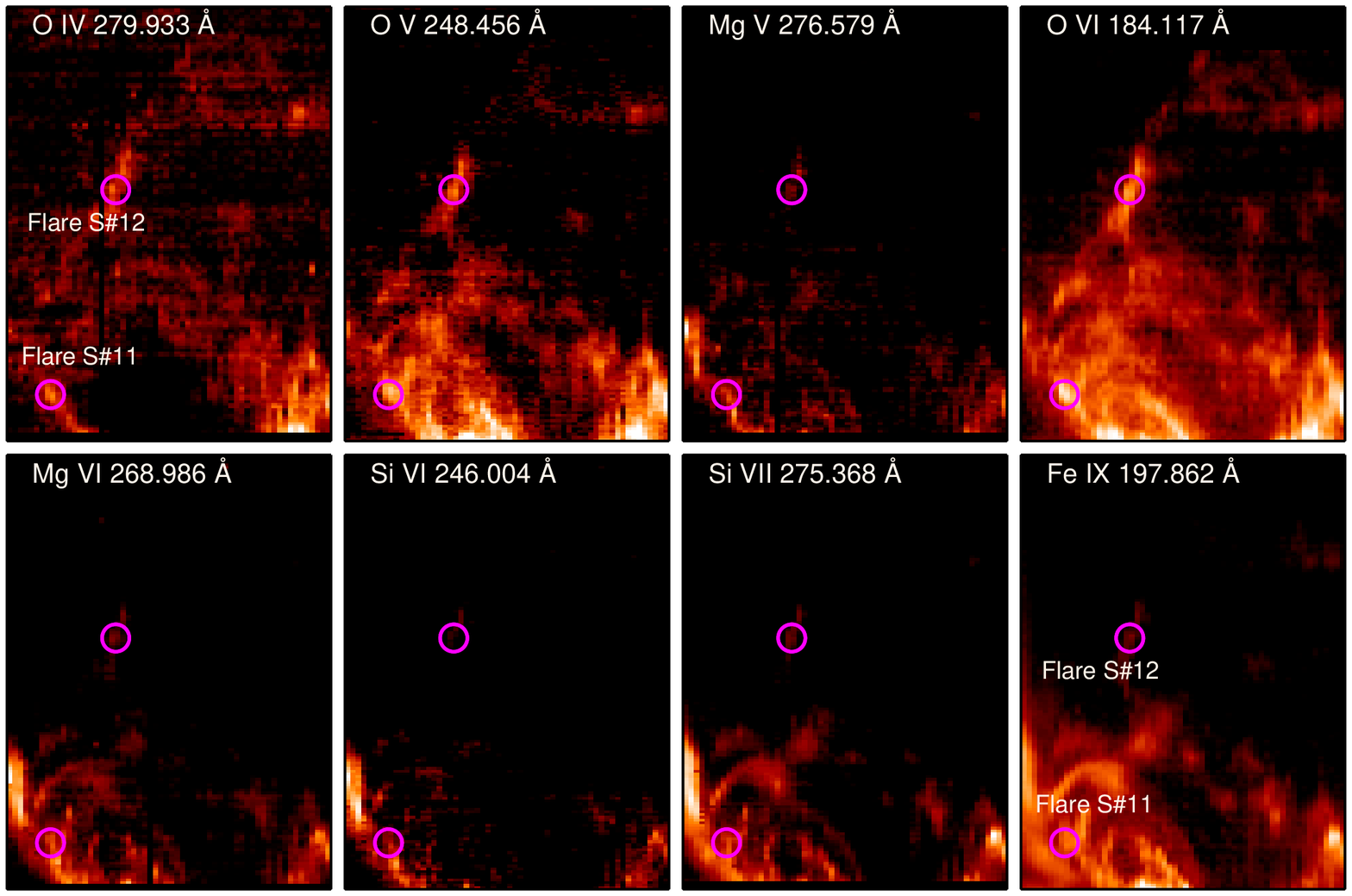}
   \includegraphics[clip,width=0.37\figurewidth]{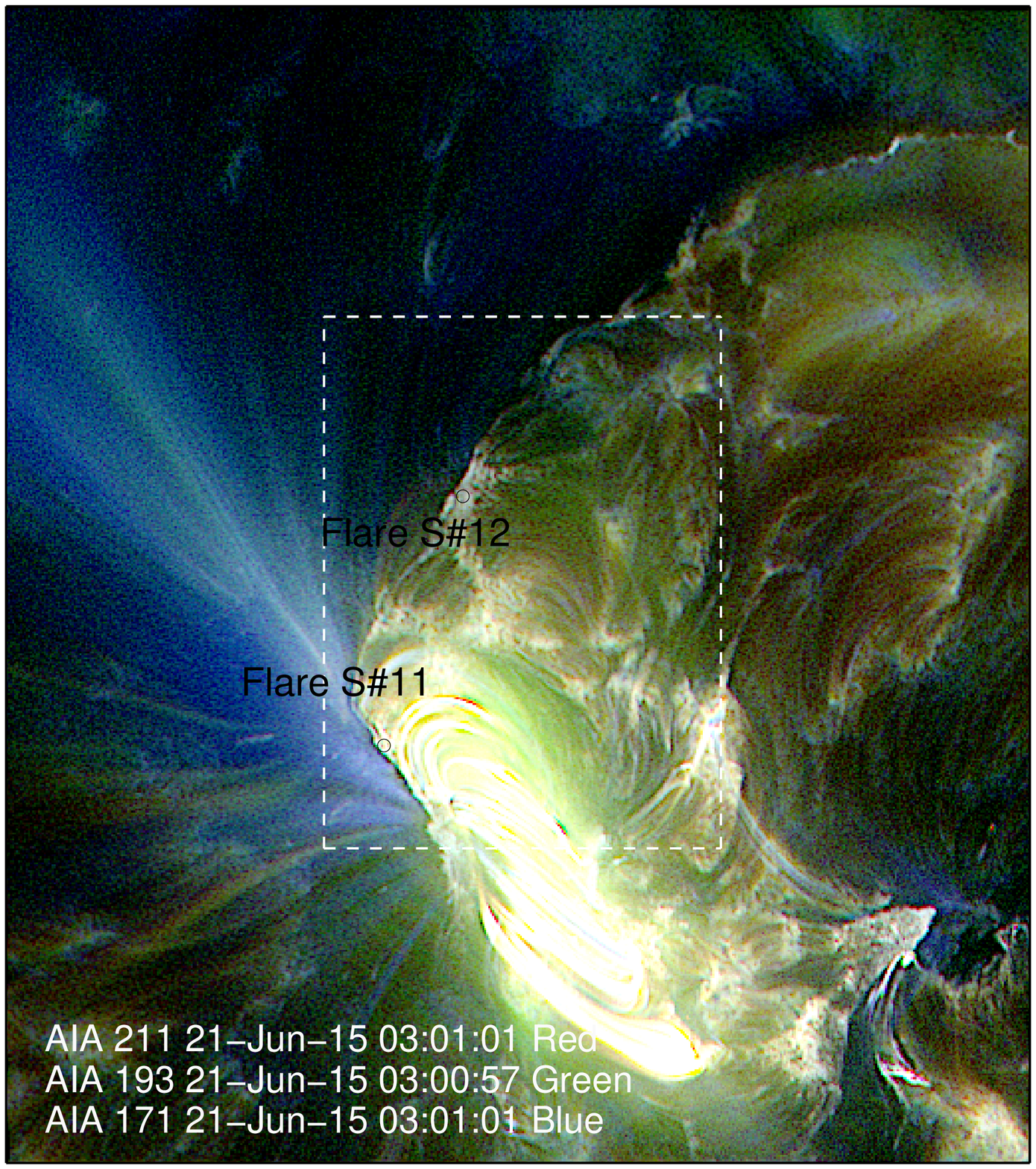}}
  \vspace{-0.1in}
  \centerline{
    \includegraphics[clip,angle=90,width=\figurewidth]{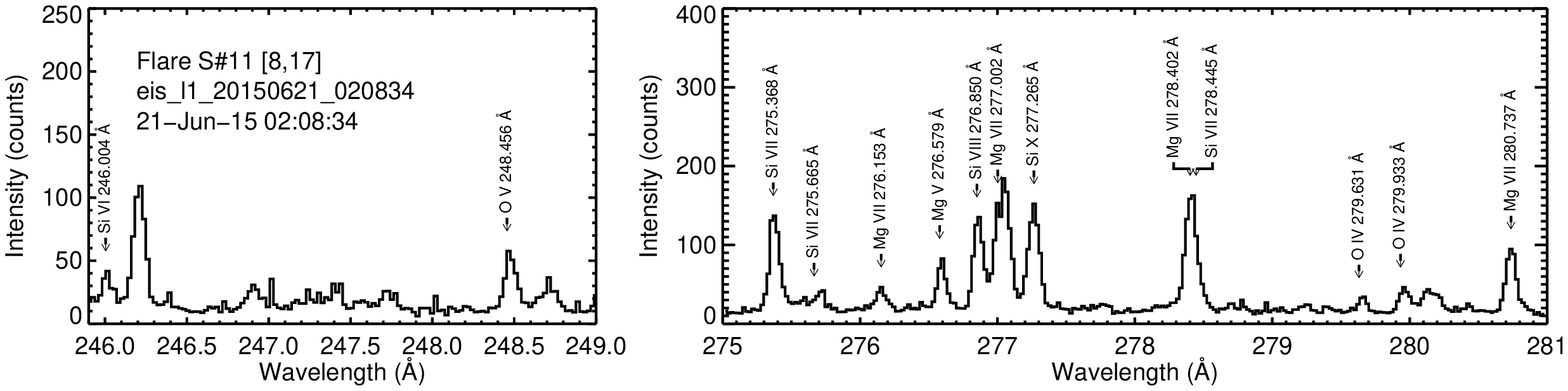}}
  \vspace{-0.2in}
  \centerline{
    \includegraphics[clip,angle=90,width=\figurewidth]{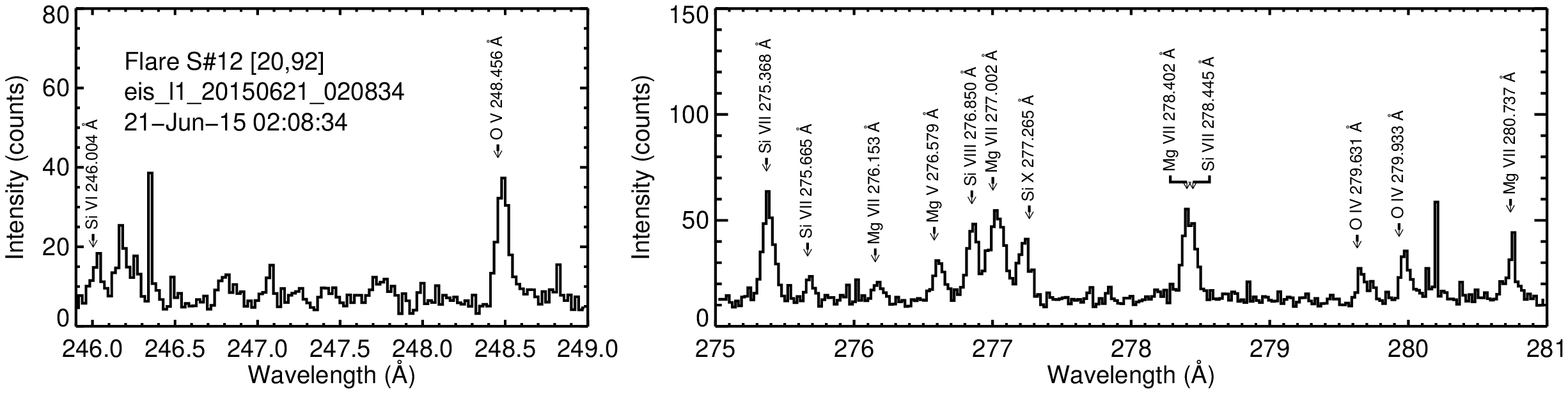}}
  \caption{EIS and AIA observations of a flaring active region taken on 2015 June 21 between
    02:08:34 and 03:09:14 UT. The format is the same as Figure~\ref{fig:r20140316}.}
\label{fig:r20150621}
\end{figure*}

\begin{figure*}[t!]
  \setlength{\figurewidth}{0.5\linewidth}  
  \centerline{
    \includegraphics[clip,angle=90,width=\figurewidth]{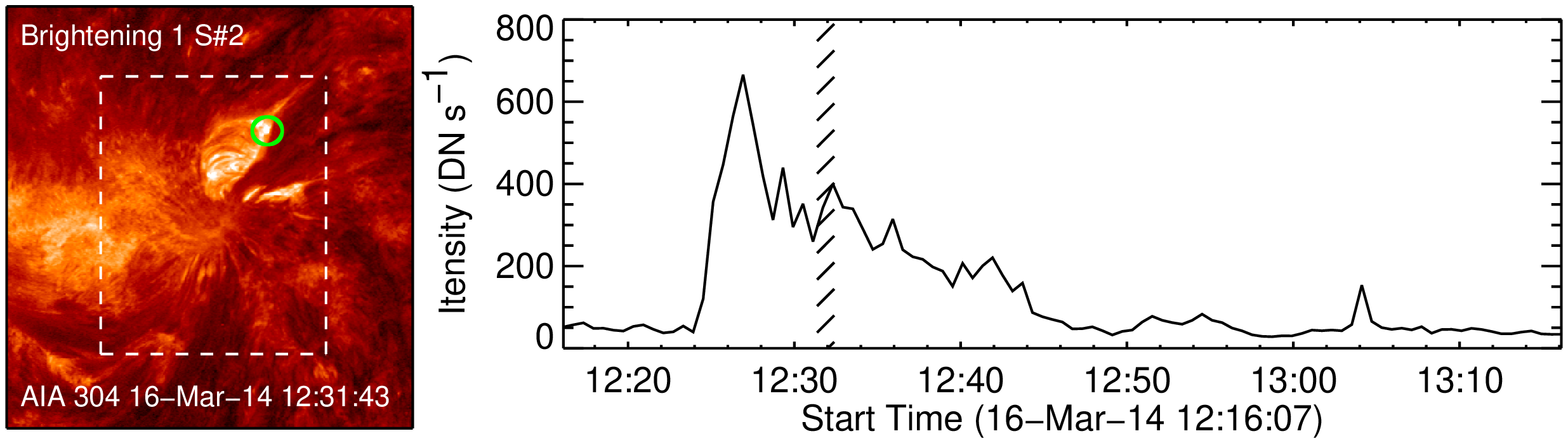}
    \includegraphics[clip,angle=90,width=\figurewidth]{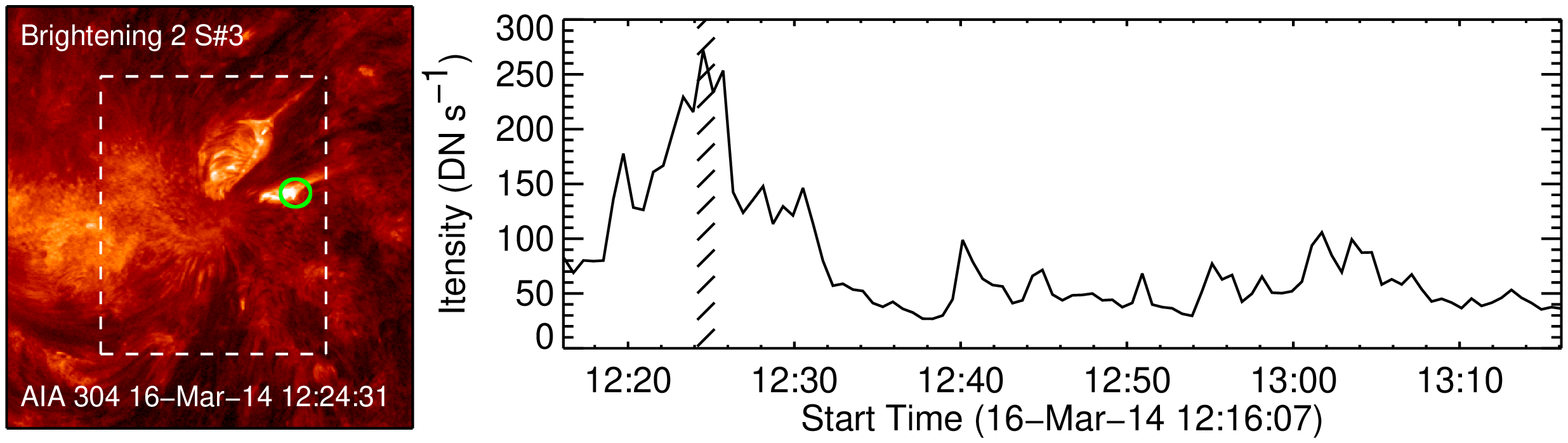}}
  \centerline{
    \includegraphics[clip,angle=90,width=\figurewidth]{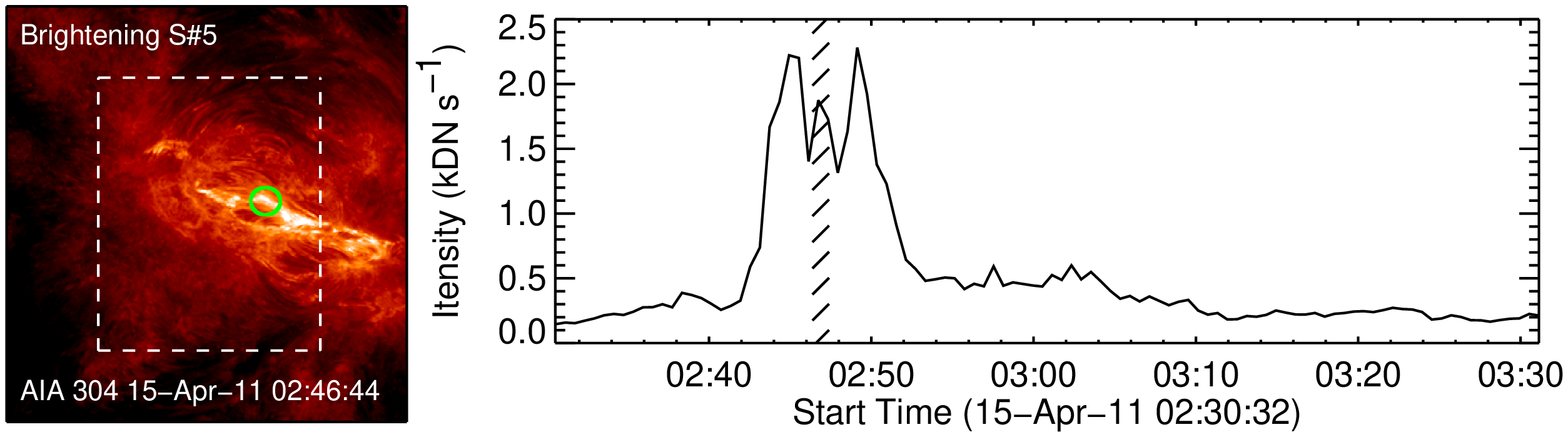}
    \includegraphics[clip,angle=90,width=\figurewidth]{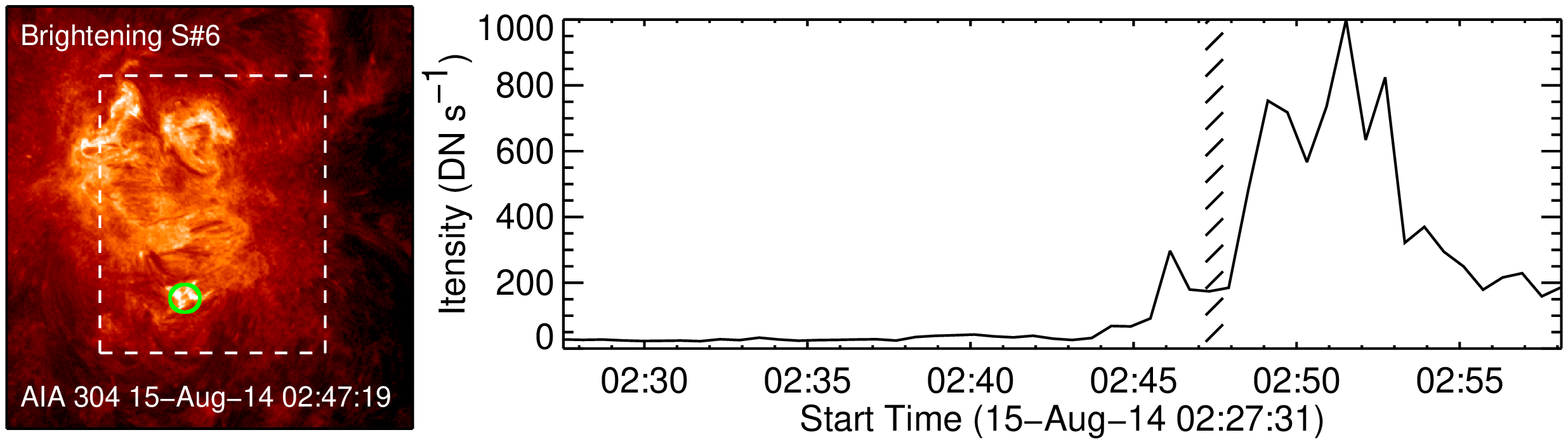}}
  \centerline{
    \includegraphics[clip,angle=90,width=\figurewidth]{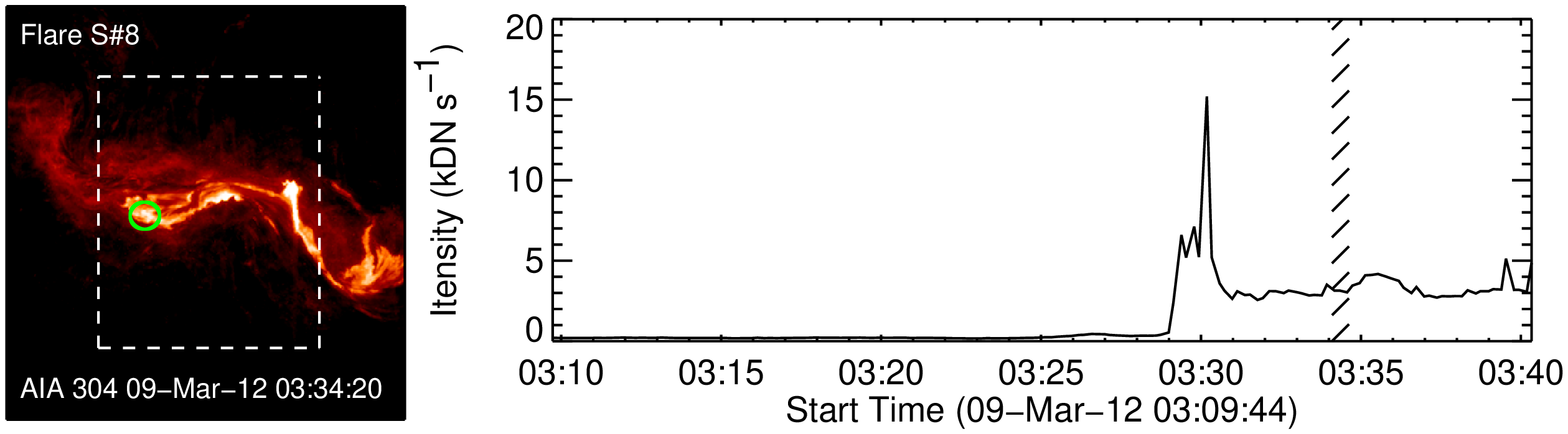}
    \includegraphics[clip,angle=90,width=\figurewidth]{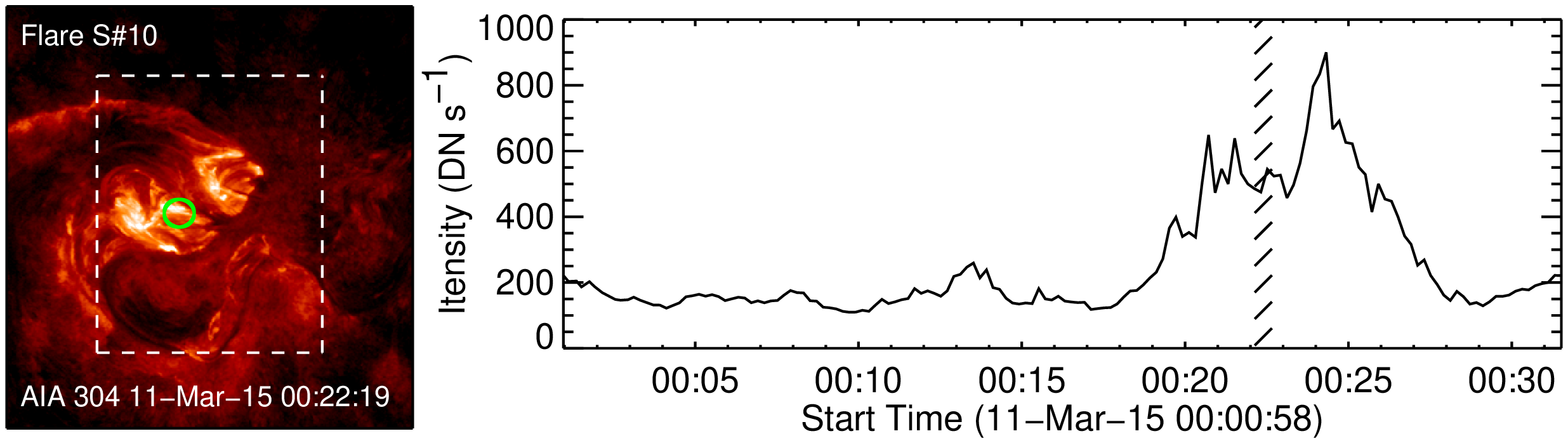}}
  \centerline{
    \includegraphics[clip,angle=90,width=\figurewidth]{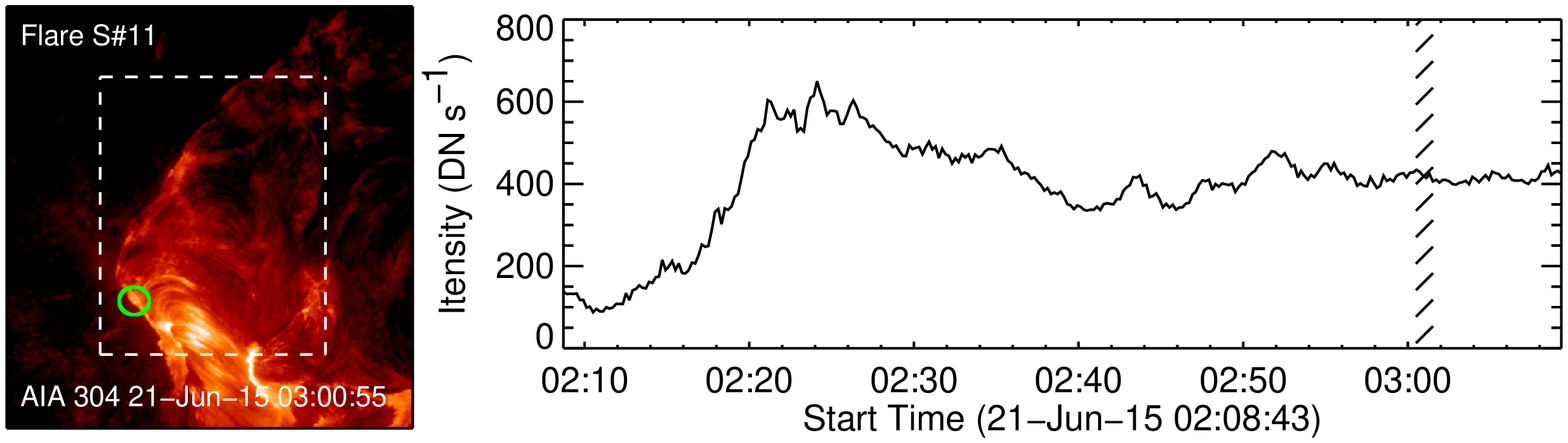}
    \includegraphics[clip,angle=90,width=\figurewidth]{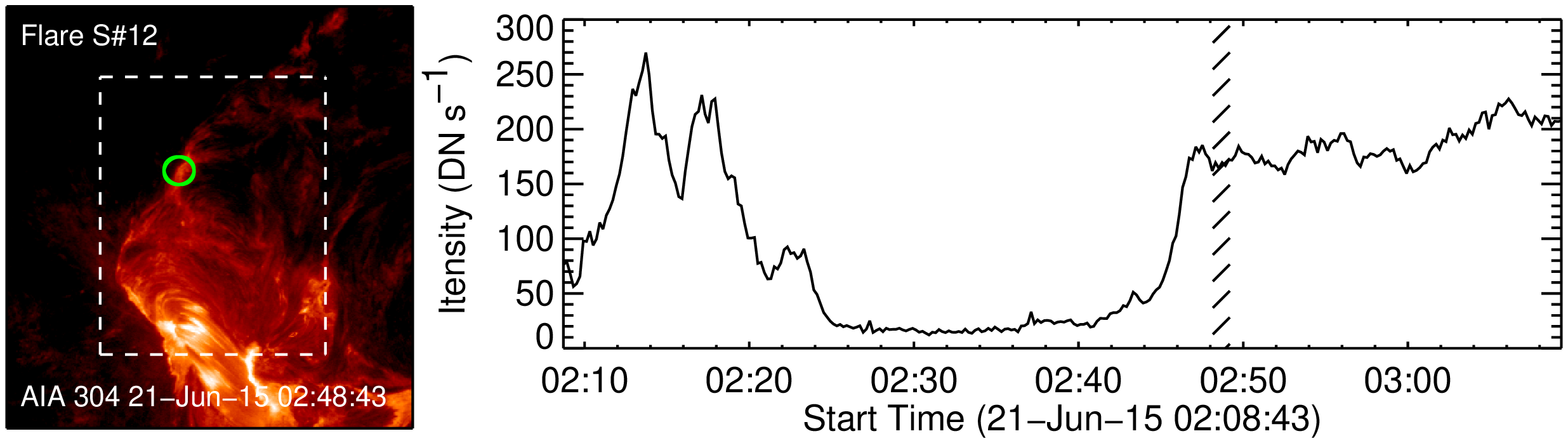}}
  \caption{AIA 304\,\AA\ intensities as a function of time for the transient heating events. The
    left panels show the AIA 304\,\AA\ image at the time of the EIS spectrum that has been
    analyzed. The position of the EIS observation is indicated by the green circle. The dashed
    line shows the field of view for the EIS raster. The hashed area on the light curve indicates
    the time for the EIS spectrum at this position. Note that the vertical scale on some of the
    light curve plots is $10^3$ DN s$^{-1}$. The light curves for the fans, which are not
    shown, are approximately constant in time. }
\label{fig:lc}
\end{figure*}

Because of telemetry constraints, EIS observations are usually structured such that a only a
subset of the full detector exposures are read out and telemetered to the ground. Since we require
a large number of emission lines for this analysis we will focus on the EIS \verb+ATLAS+ studies
that record the full wavelength range over an area of $120\arcsec\times160\arcsec$ on the sun using
the 2\arcsec\ slit and 30, 60, or 120\,s exposure times. These studies take between about 30
minutes and 2 hours to complete. Note that for these studies the slit is moved in 2\arcsec\ steps
across the sun.

These \verb+Atlas+ studies have been run over 500 times and to identify interesting features we
created a simple summary image for each observation. The summary images contain rasters for lines
formed at transition region, coronal, and flare temperatures and we manually selected observations
based on the presence of strong transition region emission. Three of the selected observations
were taken during flares as indicated by the presence of strong \ion{Fe}{24}
255.100\,\AA\ emission. Each observation is reduced in the standard way using the SSW analysis
routine \verb+eis_prep+, which removes the CCD pedestal and dark current and flags dusty and hot
pixels.  The data are not calibrated at this step but converted to ``photon events'' or counts and
are calibrated at a later stage in the processing. See \citealt{freeland1998} for a description of
the Solar Software --- SSW --- used in the analysis of these data.

In Figures~\ref{fig:r20140316}--\ref{fig:r20150621} we show EIS rasters in selected lines computed
by fitting Gaussians to the line profile of interest at every spatial pixel. For each observation
we also show an AIA three-color image computed by superimposing the 211\,\AA\ (red),
193\,\AA\ (green), and 171\,\AA\ channels (blue).

To co-align EIS to AIA we use a two step process to find the optimal spatial correlation between
the two datasets.  First we adjust the pointing provided in the EIS file headers using the results
from the SSW analysis routine \verb+eis_aia_offsets+. Second we convolve the EIS
195\,\AA\ spectral region with the AIA 193\,\AA\ effective areas and cross-correlate these
intensities along the slit with the intensities in the nearest AIA 193\,\AA\ image blurred to the
EIS spatial resolution. This process yields the co-alignment for each exposure. Note that for some
of the flare observations either the EIS or AIA observations are saturated at some positions and
the co-alignment does not work well. We exclude badly saturated exposures from consideration in
this analysis.

For each observation we have selected one to three bright features in the \ion{O}{5}
248.456\,\AA\ or \ion{Si}{6} 246.004\,\AA\ rasters and computed spectra as a function of
wavelength for these positions. These spectra account for the spatial offset of the two EIS
detectors as well as the tilt of each detector with respect to the slit. See the SSW analysis
routine \verb+eis_ccd_offset+ and the corresponding documentation for additional information.

The interpolated spectrum is computed using the data in units of counts and then converted to
physical units using both the pre-flight calibration \citep{lang2006} and the revised calibration
of \citet{warren2014b}, which includes time- and wavelength-dependent changes to the instrument
sensitivity (see \citealt{delzanna2013b} for an extensive discussion of EIS calibration
issues). For this work we use the data calibrated using the \citet{warren2014b} effective
areas. Additional comments on the calibration are provided in Appendix B.

Selected wavelength ranges from these spectra are shown in
Figures~\ref{fig:r20140316}--\ref{fig:r20150621}. There are 12 spectra in total that we will
consider here. Note that the dispersion for these spectra has not been optimized for velocity
studies and any wavelengths shifts should be interpreted as instrumental in origin. Our primary
focus here is on the interpretation of the line intensities.

We have used the co-aligned AIA observations to compute light curves for each EIS spectrum.
Unsurprisingly, the bright transition region emission identified in the EIS data generally
corresponds to either a transient heating event or a long, coronal ``fan.'' The light curves for
each of the transient brightenings are shown in Figure~\ref{fig:lc}. The light curves for the
fans, which are not shown, are remarkably constant in the AIA data.

The final step in the data reduction is to fit the emission lines of interest in each spectrum
with Gaussians so that the total line intensity can be computed. 

\section{Analysis}

As mentioned previously, the low- and high-FIP emission lines observed with EIS are not so well
matched in temperature that we can reduce the analysis to simple line ratios. Still, inspection of
the individual spectra is revealing. For the fan loops shown in Figures~\ref{fig:r20140316},
\ref{fig:r20110415}, \ref{fig:r20140815}, and \ref{fig:r20120309}, the intensity of \ion{Si}{6}
246.004\,\AA\ is always larger than that of the nearby \ion{O}{5} 248.456\,\AA\ line. Similarly,
\ion{Mg}{5} 276.579\,\AA\ is always significantly more intense than the nearby \ion{O}{4}
279.933\,\AA. In the spectra from the brightenings and the flares the opposite trend is
observed. For these spectra the intensity of \ion{O}{5} 248.456\,\AA\ is comparable to or larger
than that of \ion{Si}{6} 246.004\,\AA\ and \ion{O}{4} 279.933\,\AA\ is generally as bright or
brighter than \ion{Mg}{5} 276.579\,\AA.

The spectra suggest that there are systematic differences in the composition of the transient
events relative to the long-lived coronal fans. To put this idea on a more secure footing,
however, we must account for density and temperature effects.

As mentioned previously, almost all of the emission lines given in Table~\ref{table:atodat} have
some degree of sensitivity to density. The emissivity of \ion{O}{5} 248.456\,\AA, for example,
falls by about 25\% between $\log n = 10$ and $\log n = 11$. To account for these effects we use
the observed \ion{Fe}{13} 203.826/202.044\,\AA\ and \ion{Mg}{7} 280.737/276.153\,\AA\ line ratios
to infer the electron density. We assume that this emission is formed at the peak temperature of
formation and convert these measurements to a total pressure. The pressures derived from these
diagnostics are generally consistent to within a factor of two of each other, and we average them
to arrive at a pressure for computing the emissivities for each line of interest in each
spectrum. These measurements also follow our intuition that the pressure should be higher in
strongly heated flare loops than in the fans. In the 2012 March 9 observations, for example, the
pressure for the flare spectrum is derived to be $\log P = 16.8$ while the pressure in the fan is
observed to be $\log P = 15.9$.

To investigate the temperature structure of the observed plasma we plot the emission measure (EM)
loci curve for each line defined by
\begin{equation}
  EM(T_e) = \frac{4\pi I_{obs}}{\epsilon_\lambda(T_e, n_e)}, 
\end{equation}
where $I_{obs}$ is the observed intensity and $\epsilon_\lambda(T_e, n_e)$ is the emissivity
computed at a constant pressure. We carry out these calculations assuming both photospheric
abundances \citep{caffau2011} and our approximation to the \citet{feldman1992} abundances which
have the low-FIP elements enriched by a factor of 4 and the value for the high-FIP elements left
at their photospheric values. The EM loci curves can be interpreted as an approximate envelope to
the differential emission measure distribution, which is the best-fit solution to the ill-posed
integral equation,
\begin{equation}
  I_{obs} = \frac{1}{4\pi}\int \epsilon_{\lambda}(T_e, n_e) \xi(T_e) \, dT_e.
  \label{eq:dem}
\end{equation}
It is important to note that an EM loci curve forms an upper bound to the emission measure that
can explain the observed intensity. If one emission measure loci curve lies systematically above
another, then the intensities of both lines cannot be described by the same emission measure
distribution.

\begin{figure*}[t!]
  \centerline{
    \includegraphics[clip,angle=90,width=0.49\linewidth]{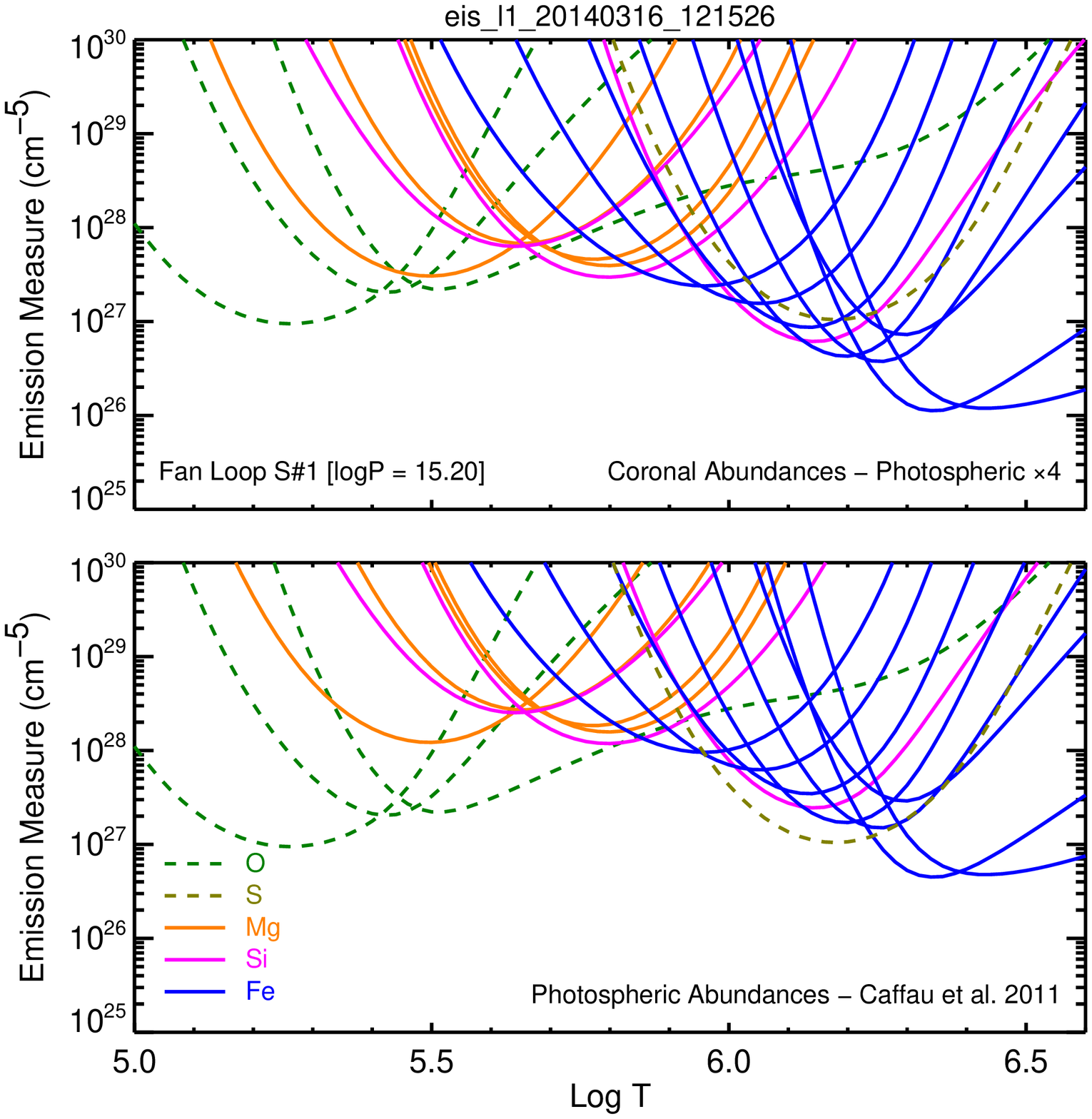}
    \hspace{0.02\linewidth}    
    \includegraphics[clip,angle=90,width=0.49\linewidth]{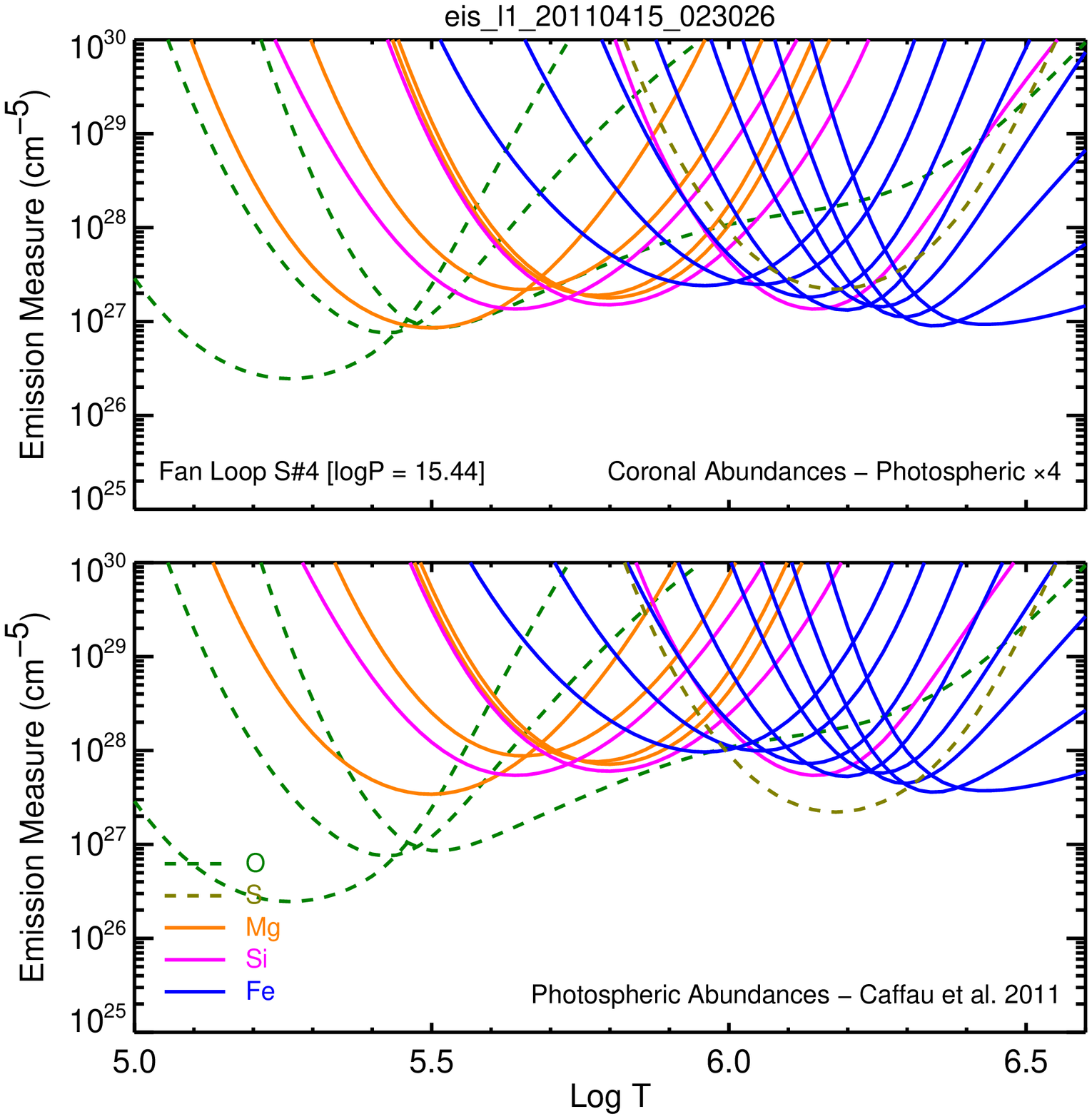}
  }
  \vspace{0.1in}  
  \centerline{
    \includegraphics[clip,angle=90,width=0.49\linewidth]{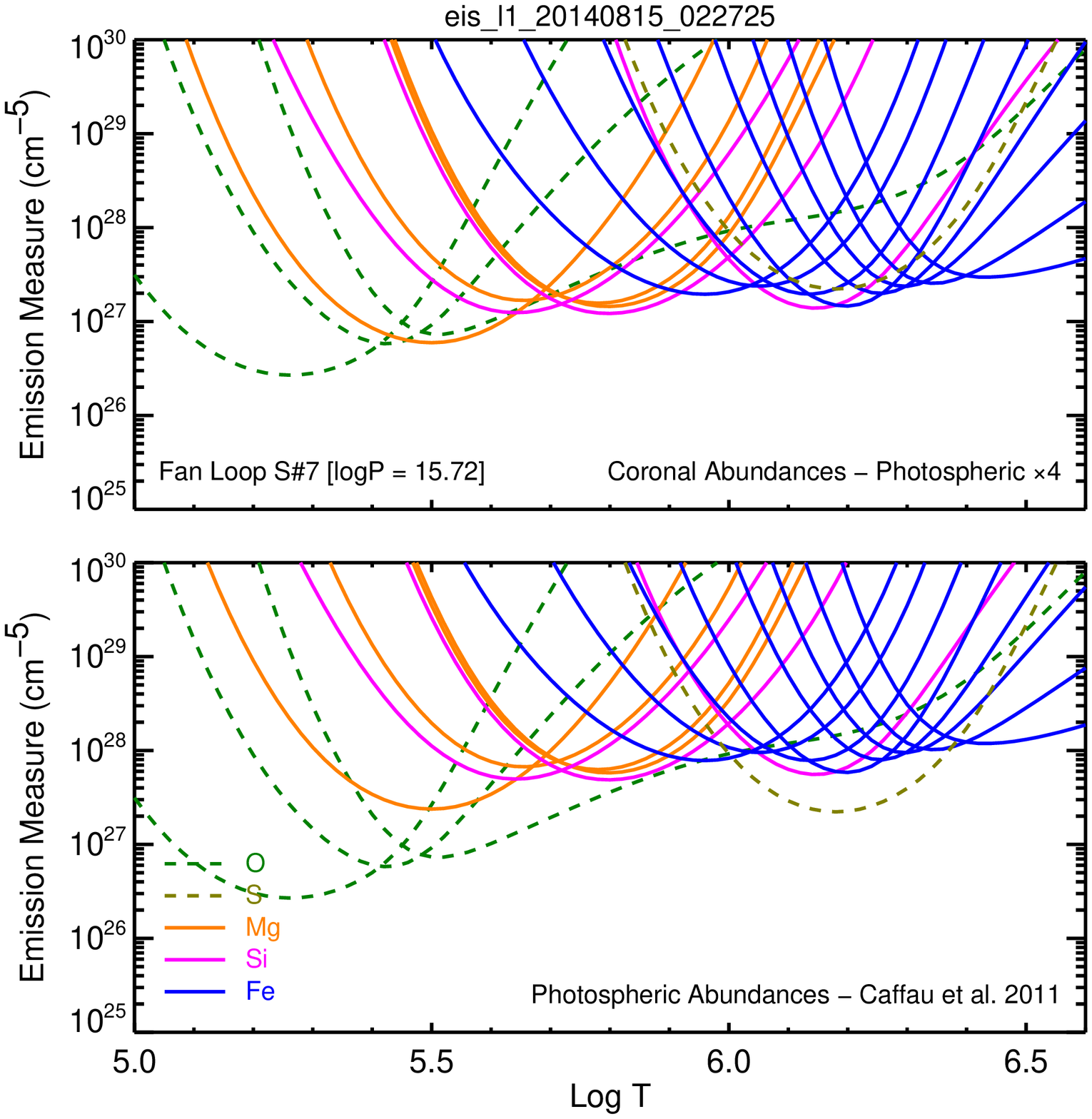}
    \hspace{0.02\linewidth}    
    \includegraphics[clip,angle=90,width=0.49\linewidth]{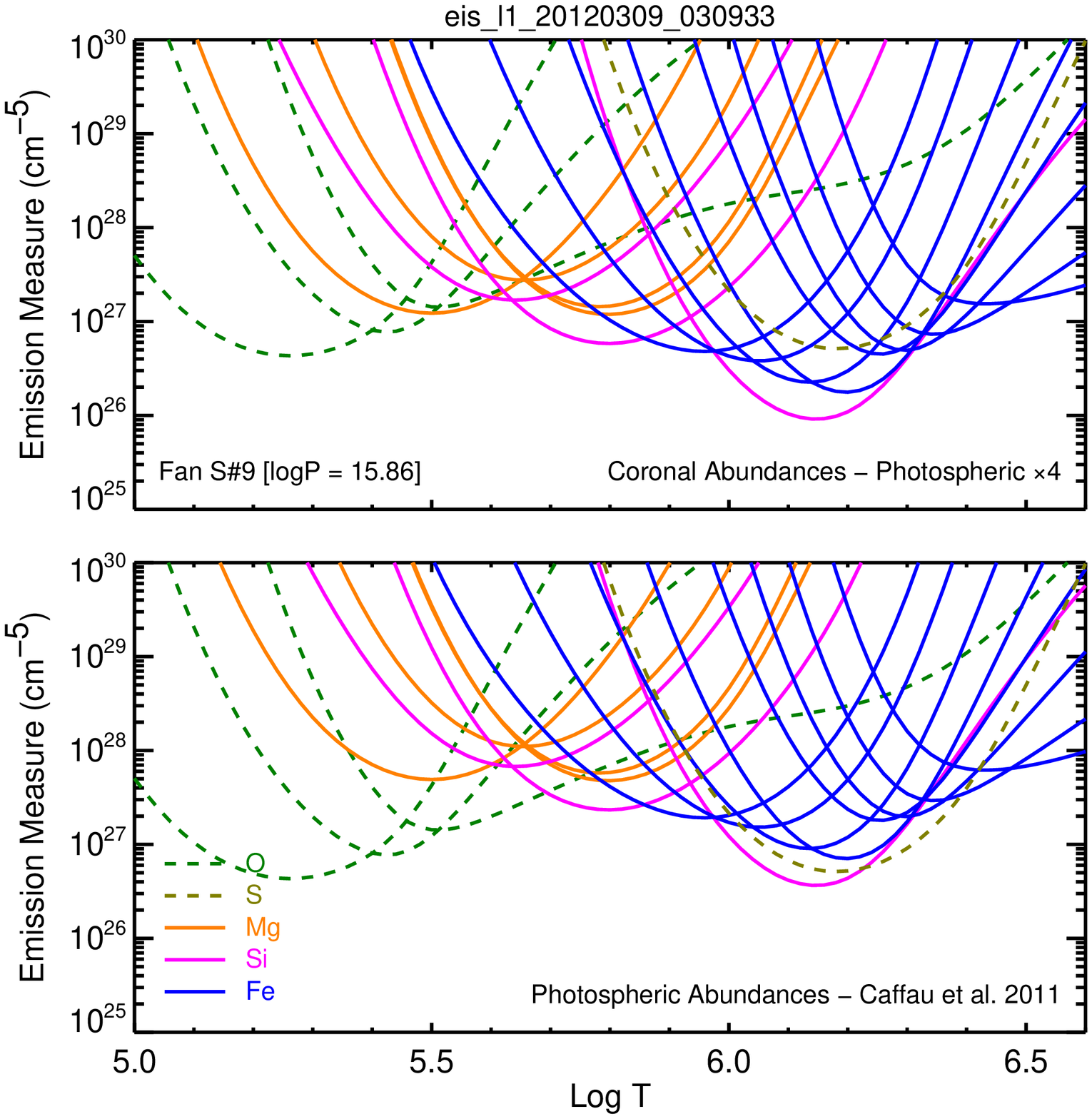}
  }
  \caption{Emission measure loci curves for the four fan observations. For each observation
    calculations for both photospheric and coronal abundances are shown. The observed intensities
    of the transition region O and Mg lines are more consistent with a coronal composition. The
    pressure derived from the \ion{Fe}{13} and \ion{Mg}{7} densities are indicated on each set of
    plots. }
  \label{fig:loci_fans}
\end{figure*}

\begin{figure*}[t!]
  \centerline{
    \includegraphics[clip,angle=90,width=0.49\linewidth]{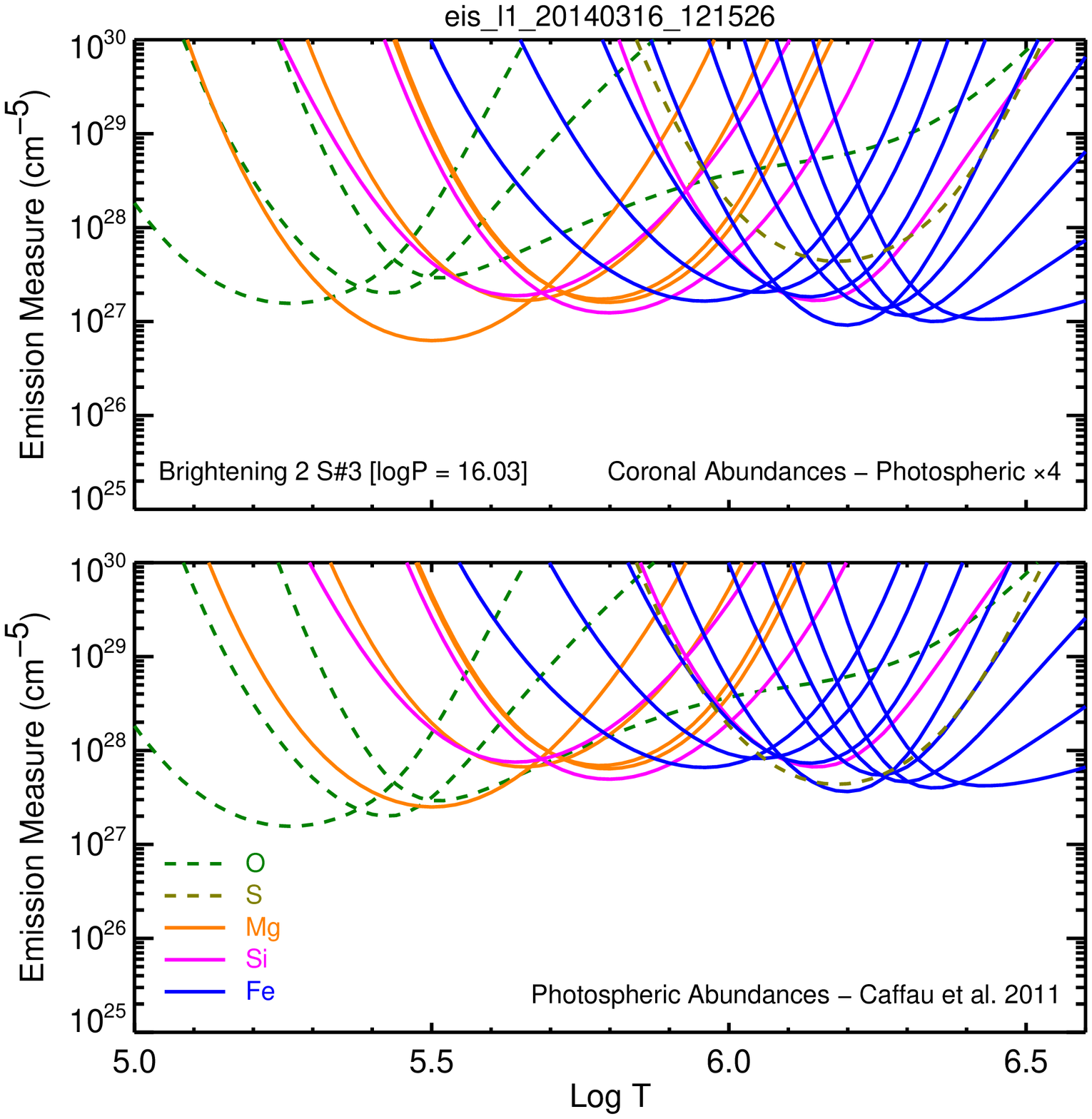}
    \hspace{0.02\linewidth}
    \includegraphics[clip,angle=90,width=0.49\linewidth]{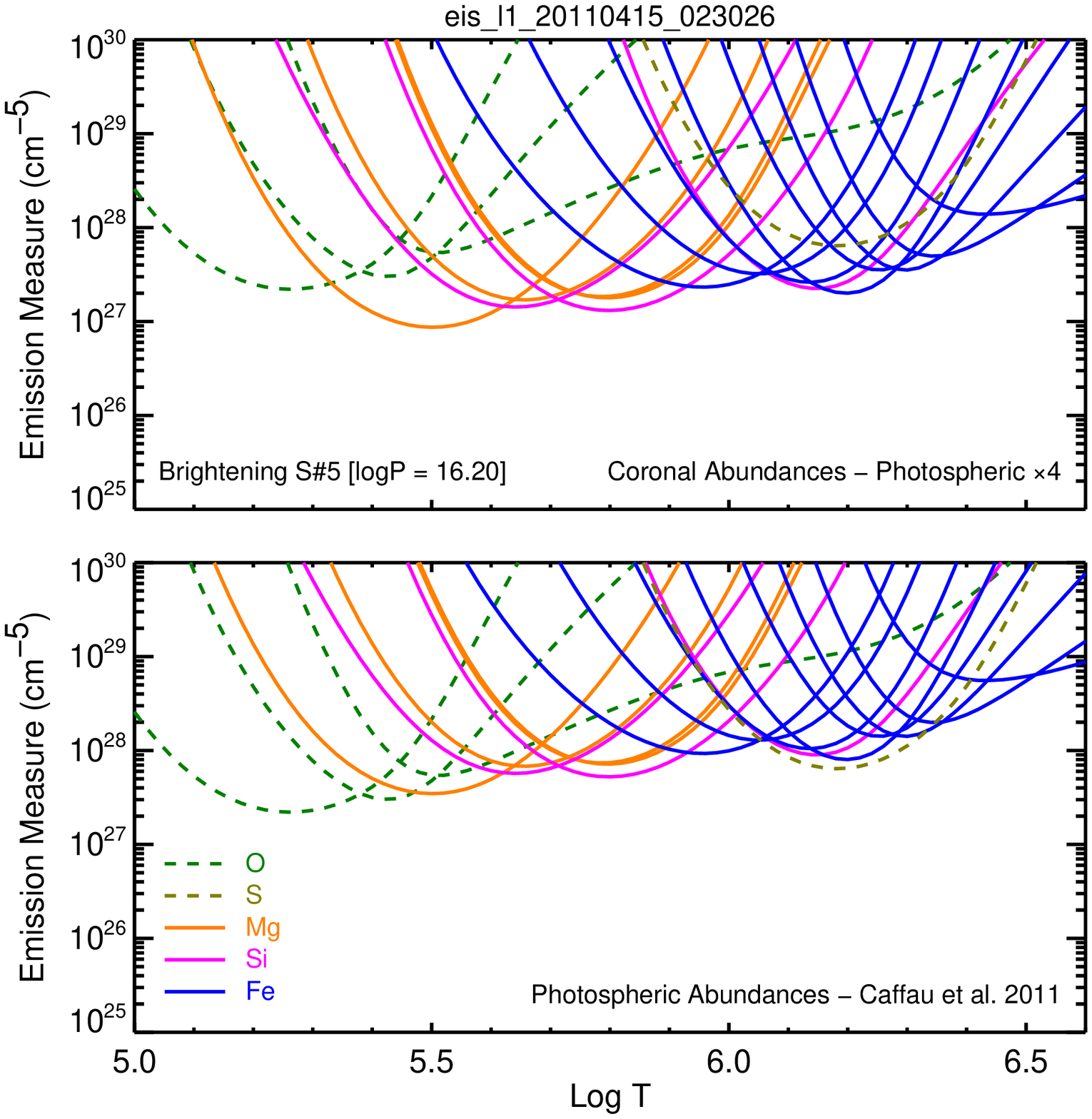}
  }
  \vspace{0.1in}
  \centerline{
    \includegraphics[clip,angle=90,width=0.49\linewidth]{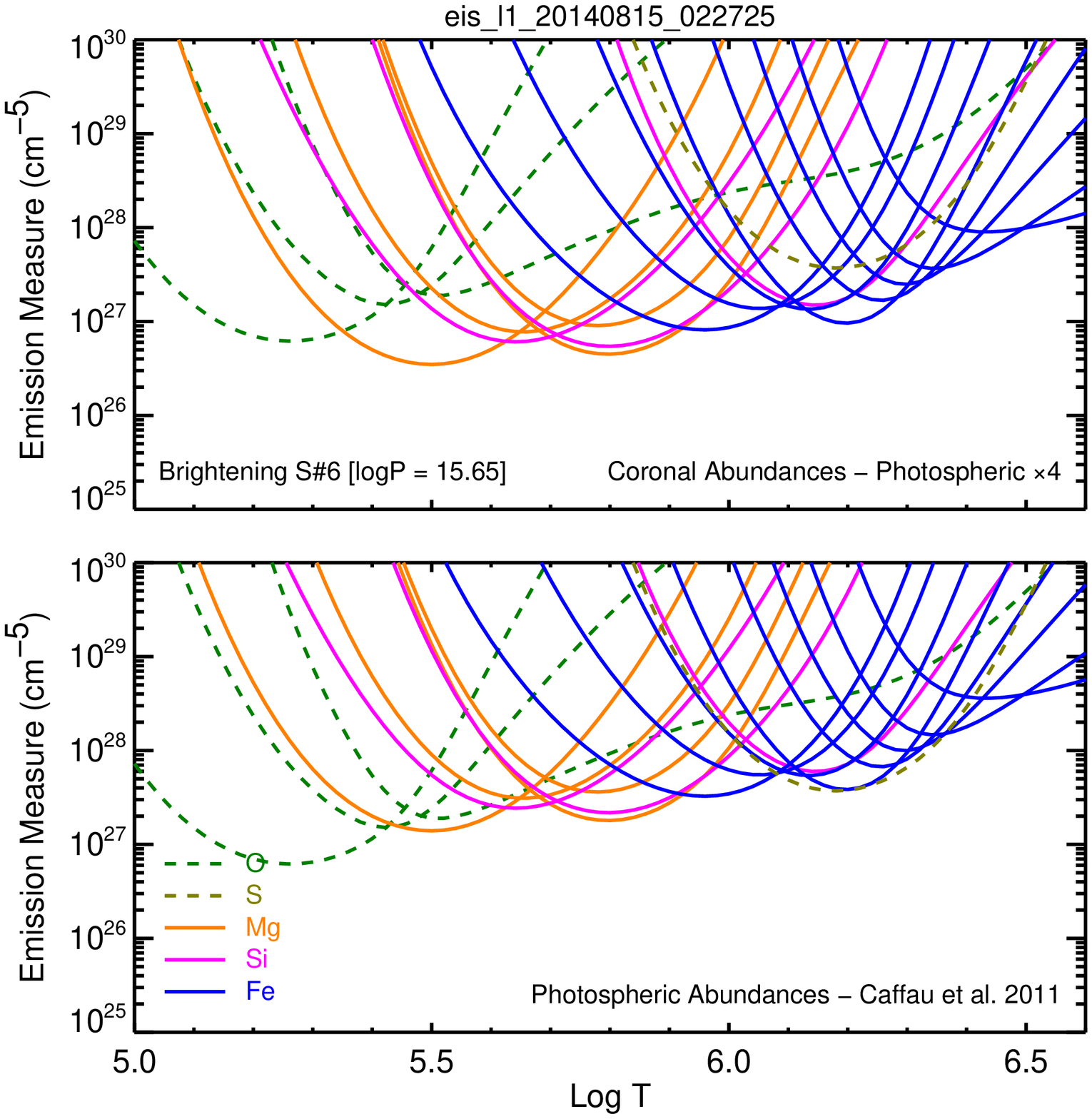}
    \hspace{0.02\linewidth}    
    \includegraphics[clip,angle=90,width=0.49\linewidth]{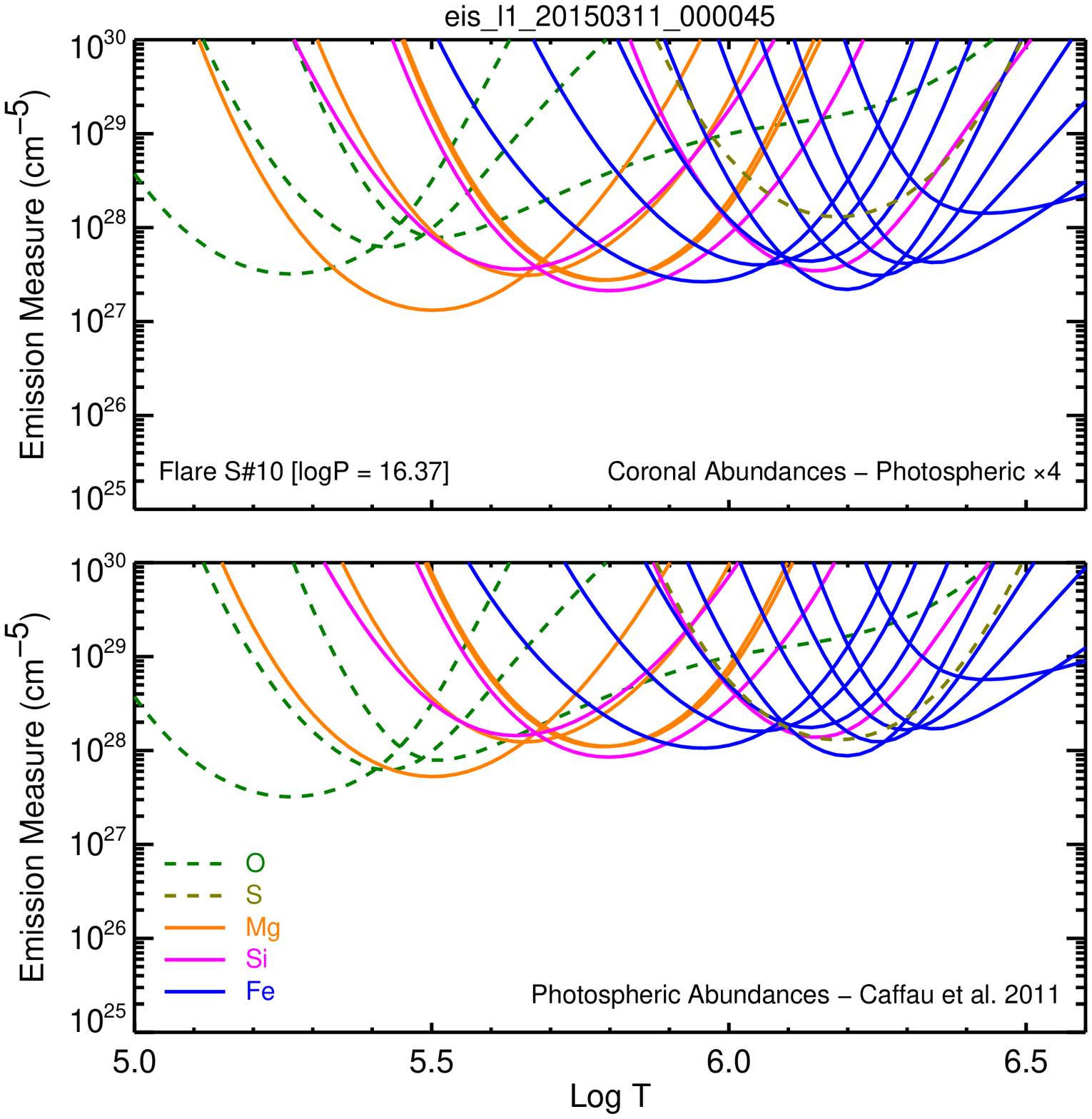}
  }
  \caption{Emission measure loci curves for four of the transient heating events. The format is
    the same as Figure~\ref{fig:loci_fans}. For these features the observed intensities of the
    transition region O and Mg lines are more consistent with a photospheric composition. The
    emission measure loci for the other events are similar to these. }
  \label{fig:loci_flares}
\end{figure*}

Emission measure loci plots for many of the observations are shown in Figures~\ref{fig:loci_fans}
and \ref{fig:loci_flares}. These plots are consistent with the simple inspection of the
spectra. In Figure~\ref{fig:loci_fans} we show the EM loci for all of the observations of the
fans. For each case we see that the observed intensities for the low- and high-FIP elements are
more consistent with a coronal composition than with a photospheric one. In each case the
intensities of the O lines cannot be reconciled with the intensities of the \ion{Mg}{5} and
\ion{Mg}{6} lines assuming a photospheric composition. 

The EM loci plots for the transient heating events show the opposite trend. For these spectra we
see that the intensities for the low- and high-FIP elements are more consistent with a
photospheric composition than with a coronal one. In these cases the intensities of the O lines
cannot be reconciled with the intensities of the \ion{Mg}{5} and \ion{Mg}{6} lines assuming a
coronal composition.

As a check on these trends we have also included EM loci curves for the \ion{S}{10}
264.233\,\AA\ line. S has an FIP of 10.4\,eV and sits on the boundary between low- and high-FIP
elements. Nevertheless, it has proven to be a useful diagnostic in studies of coronal outflows
\citep[e.g,][]{brooks2015, brooks2011} and in active regions
\citep[e.g.,][]{baker2015,delzanna2013}. The observations of this coronal line are generally
consistent with the interpretation of the transition region emission. The EM loci for \ion{S}{10}
is generally closer to those of the coronal Fe and Si lines assuming a coronal composition for the
fans. For the transient heating events the observed S emission is generally more consistent with a
photospheric composition, although there are some exceptions. We show EM loci plots for only 4 of
the 8 spectra for the transient heating events. The remaining plots are very similar to what is
displayed here. 

As a final check on the interpretation of the spectra we have computed differential emission
measure distributions for each spectrum assuming both coronal and photospheric abundances for the
emissivities. We used the Monte Carlo Markov Chain (MCMC) emission measure algorithm
\citep{kashyap1998,kashyap2000} distributed with the \verb+PINTofALE+ spectral analysis package to
solve Equation~\ref{eq:dem}. This algorithm has the advantage of not assuming a functional form
for the differential emission measure. The MCMC algorithm also provides for estimates of the error
in the EM by calculating the emission measure using perturbed values for the intensities.  These
calculations are also consistent with our interpretation of the spectra and the EM loci
curves. Again, for the fans the assumption of coronal abundances leads to smaller $\chi^2$ values
for the modeled intensities. For the heating events the assumption of photospheric abundances
leads to smaller $\chi^2$ values.

The DEM calculations also point to the limitations of the EIS data for exploring this issue. For
example, since there is no low-FIP line that has good overlap in temperature with \ion{O}{4} the
DEM can be adjusted to match its observed intensity regardless of the assumed abundance. Also,
since the emissivity of \ion{O}{6} has been adjusted it is useful to perform the DEM calculation
without it. In this case a reasonable fit to the fan observations can be achieved for photospheric
abundances if the DEM increases very sharply at around $\log T = 5.5$. For the brightenings
\ion{O}{6} plays less of a role in determining the structure of the DEM.  In these spectra
\ion{Mg}{5} forms a lower bound to the DEM that is inconsistent with the observed \ion{O}{4} and
\ion{O}{5} emission if coronal abundances are assumed. These conclusions are evident from
inspection of the EM loci curves and we have chosen not to display the DEM curves.

\section{Summary and Discussion}

We have presented the analysis of EIS spectra observed during transient heating events. During
these times the observed transition region emission is consistent with a photospheric
composition. This is inferred from the strong intensities in \ion{O}{4}--\ion{O}{6} emission lines
relative to Mg and Si lines formed at similar temperatures. The relative intensities for these
lines in the fan loops are much different. In these locations the Mg and Si lines are strong
relative to those from  \ion{O}{4}--\ion{O}{6}. These trends are clearly reflected in the emission
measure loci plots that describe the temperature structure of the plasma at these locations.

Our analysis suggests that the plasma composition could be an important signature of the coronal
heating process. Plasma that is evaporated from the chromosphere by strong, impulsive heating,
such as in a flare or jet may not show any enrichment in low-FIP elements. Coronal abundances may
be a feature of plasma confined to long-lived structures, such as in the fan loops or the
high-temperature emission in active region cores.

It is interesting to interpret these results in the context of abundance measurements at
transition region temperatures in the quiet Sun. Analysis by \citet{laming1995} and
\citet{young2005} have indicated that the FIP effect is greatly reduced or even absent in the
quiet Sun at transition region temperatures. In Appendix B we present some EIS quiet Sun spectra
which also show a strong reduction in the FIP effect at transition region temperatures.  These
observations could be consistent with the idea that transition region emission is dominated by
short-lived structures, the unresolved fine structures of \citet{feldman1983}, for which the
abundances are predominately photospheric.

Unfortunately, a direct connection between the coronal heating process and abundance variations is
only a conjecture at this point. More analysis is needed to fully understand how the composition
varies from feature to feature and with time. Ideal observations would include many low- and
high-FIP emission lines, very high spatial resolution, and high cadence. The Interface Region
Imaging Spectrograph (IRIS, \citealt{depontieu2014}) can achieve the high spatial resolution and
cadence needed to study the morphology and dynamics of transition region
structures. \citet{hansteen2014}, for example, have used IRIS observations to find evidence for
highly variable, low lying loops at transition region temperatures. It is not clear, however, that
IRIS observes enough low and high-FIP emission lines to infer the plasma composition for these
structures. It is possible that coordinated observations between IRIS and EIS will provide useful
information, but the mismatch in spatial resolution could be difficult to overcome. Progress on
understanding abundance variations and their connection to coronal heating may need to wait for
future instrumentation, such as the Marshall Grazing Incidence X-ray Spectrograph (MaGIXS,
\citealt{kobayashi2011}) or the EUV spectrograph proposed for \textit{Solar-C}
\citep{teriaca2012}.


\acknowledgments Hinode is a Japanese mission developed and launched by ISAS/JAXA, with NAOJ as
domestic partner and NASA and STFC (UK) as international partners. It is operated by these
agencies in co-operation with ESA and NSC (Norway). The authors would like to acknowledge many
helpful conversations on this work with Peter Young. The authors would also like to thank Remy
Freire, who helped develop the EIS-AIA co-alignment software and began the analysis of several of
these events during a summer internship at NRL. This work has been sponsored by NASA's
\textit{Hinode} project.

\appendix
\section{\ion{O}{6} Intensities in the Quiet Sun}

\begin{figure}[t!]
\centerline{\includegraphics[clip,angle=90,width=\linewidth]{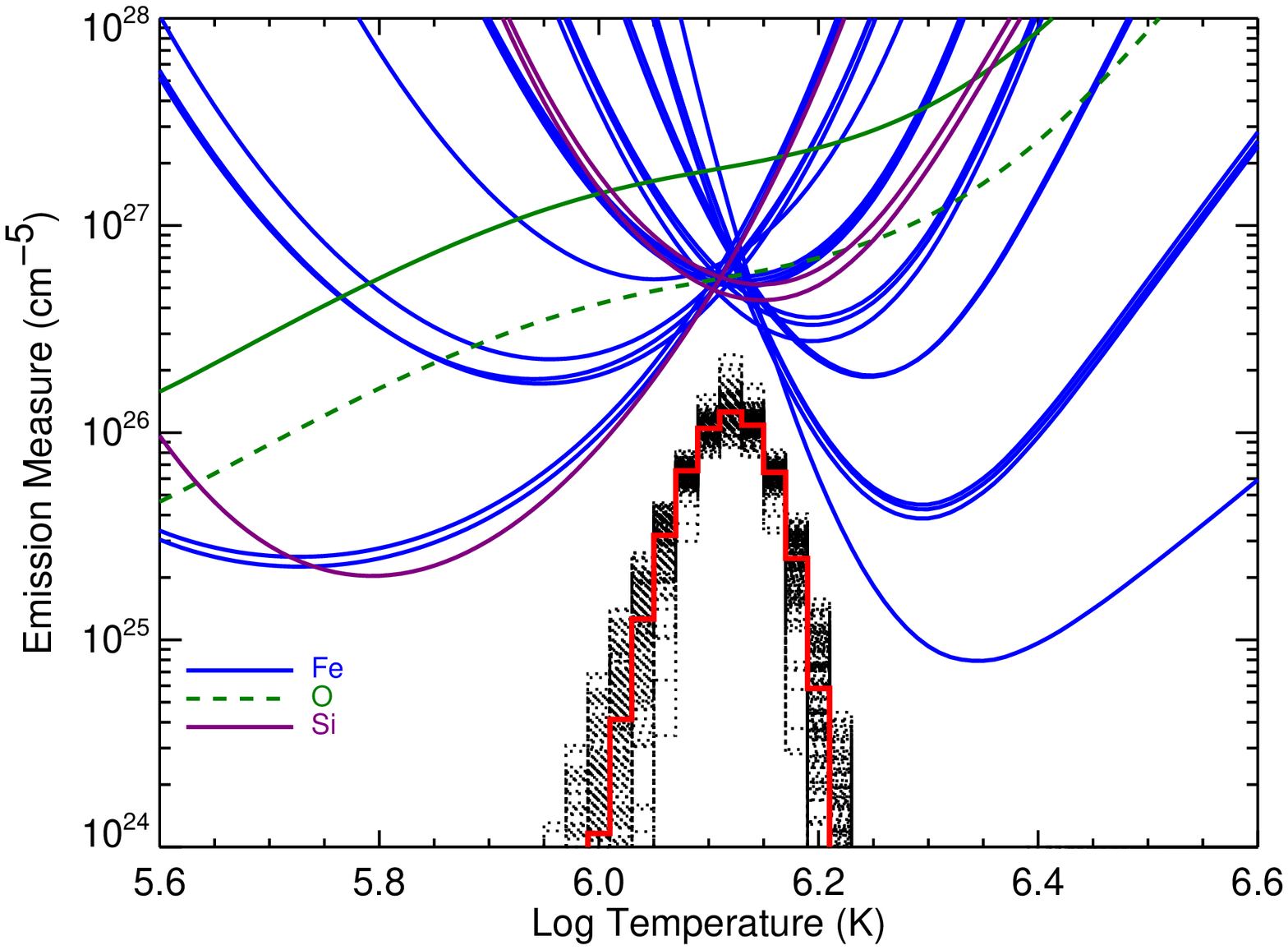}}
 \caption{The emission measure for the quiet Sun, off-limb observations. The red line is the
 differential emission measure as a function of temperature. Emission measure loci curves are
 shown for the lines given in Table~\ref{table:offlimb}. The EM loci for \ion{O}{6} 184.117\,\AA\ is
 shown scaled (green dash) and unscaled (green solid). }
\label{fig:offlimb}
\end{figure}

\begin{figure*}[b]
\centerline{
    \includegraphics[clip,angle=90,width=\linewidth]{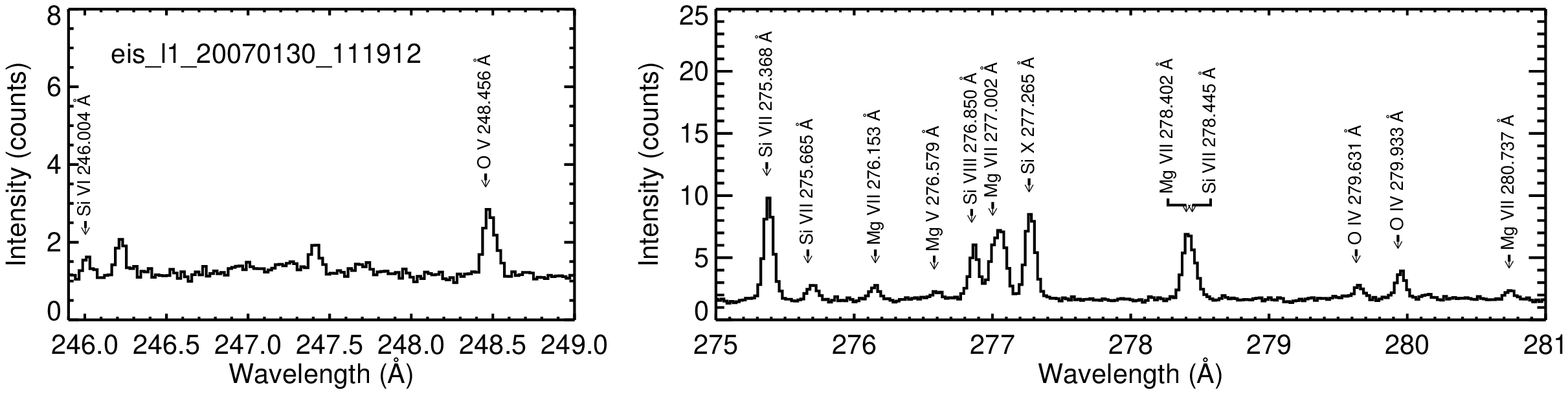}
 }
\vspace{-0.2in}     
\centerline{        
   \includegraphics[clip,angle=90,width=\linewidth]{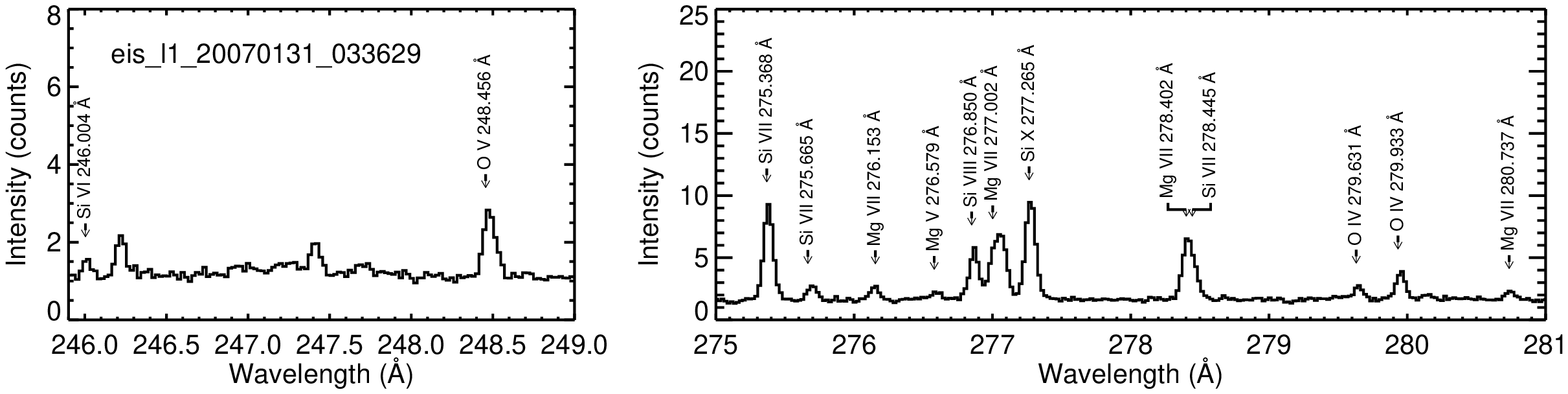}
}
\caption{Spatially and temporally averaged EIS spectra from on-disk, quiet Sun observations. These
    spectra show that the high-FIP O lines are strong relative to the low-FIP Mg and Si lines,
    suggesting a composition close to that of the photosphere for transition region emission in
    the quiet Sun. } \label{fig:spec_disk}
\end{figure*}

\begin{figure*}[t!]
  \centerline{
    \includegraphics[clip,angle=90,width=0.49\linewidth]{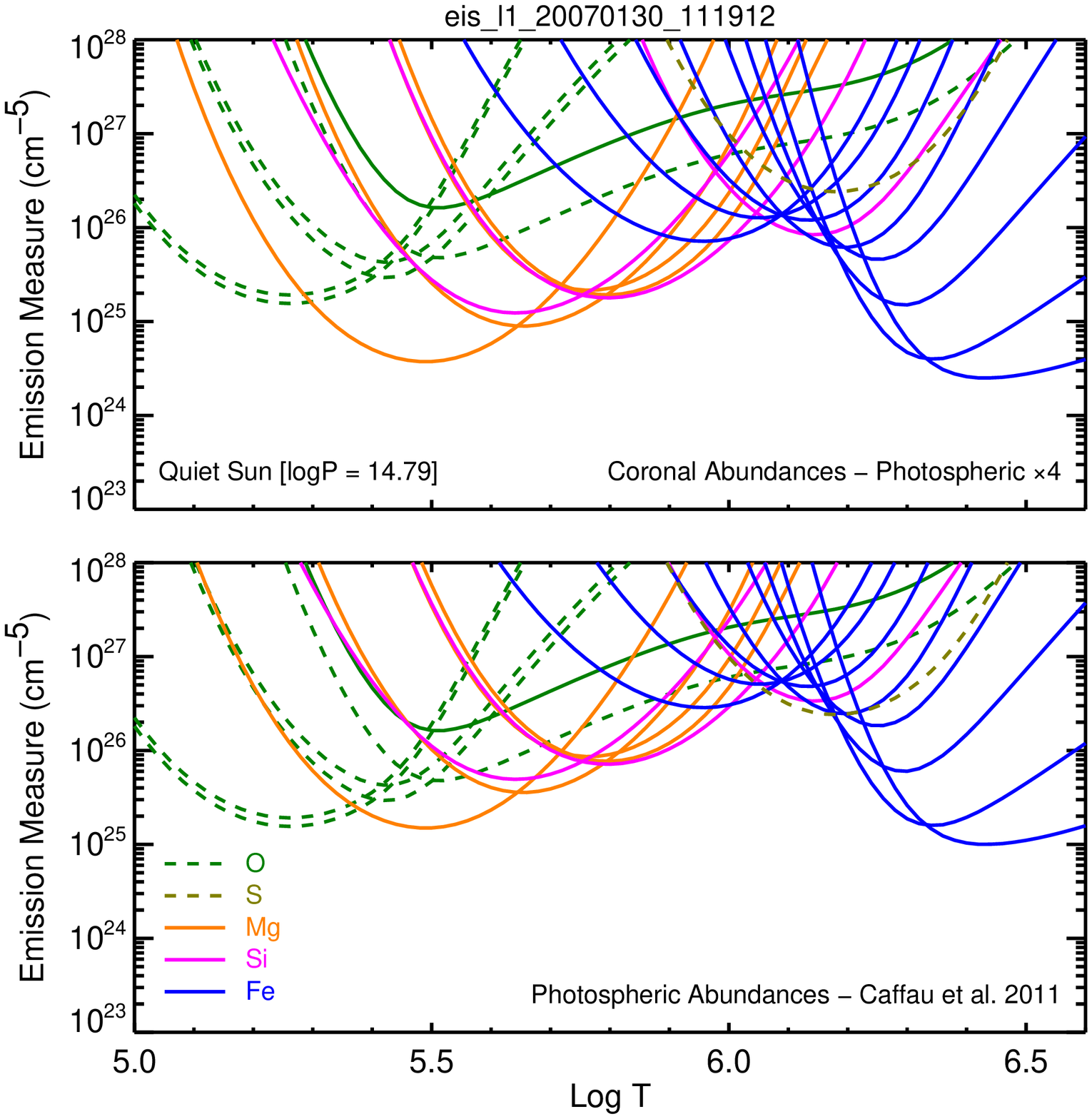}
    \hspace{0.02\linewidth}    
    \includegraphics[clip,angle=90,width=0.49\linewidth]{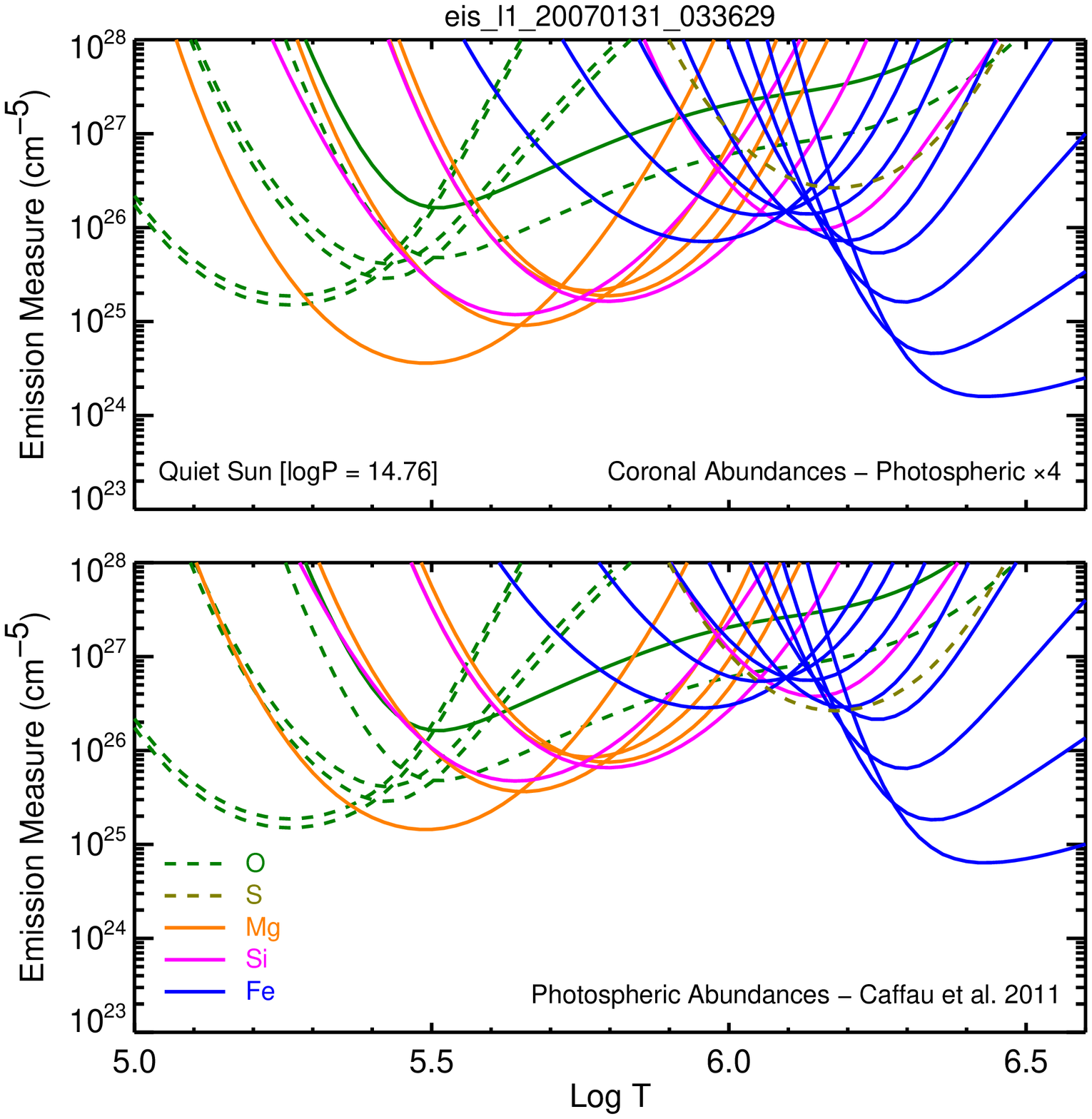}
  }
\caption{Emission measure loci curves for the on-disk, quiet Sun spectra shown in
  Figure~\ref{fig:spec_disk}. The format is the same as Figures~\ref{fig:loci_fans}
  and \ref{fig:loci_flares}. The quiet Sun intensities appear to be more consistent with
  photospheric abundances than with coronal abundances. In either case the \ion{O}{6}
  184.117\,\AA\ line is discrepant. Multiplying it by the factor of 3.4 derived from the off-limb
  spectra brings it into closer agreement with the other O lines. the solid green line is the
  emissivity as calculated from CHIANTI.}  \label{fig:loci_disk}
\end{figure*}

In this section we discuss the intensity of the \ion{O}{6} 184.117\,\AA\ line. Observations
from \ion{O}{6} are needed to improve the overlap between the low- and high-FIP lines in the EIS
spectra. In our analysis we have scaled it by a factor of 3.4 to bring it into agreement with the
other emission lines formed at similar temperatures. Here we discuss the supporting evidence for
this adjustment.

One way to examine the relative consistency of spectral data is to consider observations from the
quiet Sun above the limb. Numerous observations have shown that the off-limb, quiet corona has a
very narrow temperature
distribution \citep[e.g.,][]{raymond1997,feldman1998,landi2002,warren2009}. This makes it easy to
identify possible inconsistencies between the atomic data and the observed intensities.

To illustrate this type of analysis we use the 2007 November 4 off-limb observations presented
by \citet{warren2014b}. These spectra were formed by averaging about 38 300\,s exposures
together. The observed intensities for the lines relevant to our study are shown in
Table~\ref{table:offlimb}. These intensities use the adjusted calibration
of \citet{warren2014b}. In Figure~\ref{fig:offlimb} we show the emission measure loci and
differential emission measure distribution derived from these intensities. The Fe and Si lines
from ions formed relatively near $\log T = 6.1$ all intersect close to a point, suggesting
nearly isothermal plasma. The DEM is also relatively narrow with a Gaussian width of about
$10^5$\,K. Here we have assumed that the DEM is Gaussian in shape. A similar result is obtained
using MCMC. When compared with the other lines used in this analysis, the intensity
of \ion{O}{6} 184.117\,\AA\ is too high by a factor of about 3.4. This is indicated by the EM loci
curves shown in Figure~\ref{fig:offlimb}.

Some previous studies have identified problems in the analysis of the emission from Li-like ions
such as \ion{O}{6} \citep[e.g.,][]{dupree1972,warren2005}. However, many previous analyses of off
limb spectra have successfully used the longer wavelength \ion{O}{6} lines to measure the
composition of the corona \citep[e.g.,][]{feldman1998}.  \citet{muglach2010} compared the long
wavelength \ion{O}{6} lines observed with SUMER and the short wavelength \ion{O}{6} lines observed
with EIS and found that their intensities could not be reconciled. 

\begin{deluxetable}{rrrr}
  \tablewidth{2.5in}
  \tabletypesize{\scriptsize}  
  \tablecaption{Modeling EIS Off-Limb Observations\tablenotemark{a}}
  \tablehead{
    \multicolumn{1}{c}{Line} &
    \multicolumn{1}{c}{$I_{obs}$} &
    \multicolumn{1}{c}{$I_{dem}$} &
    \multicolumn{1}{c}{$R$}
    }
  \startdata
      \ion{Fe}{8} 185.213    &     21.30    &     21.92    &       1.0   \\
      \ion{Fe}{8} 186.601    &     17.45    &     16.08    &       1.1   \\
      \ion{Si}{7} 275.368    &      8.39    &      8.96    &       0.9   \\      
      \ion{Fe}{9} 188.497    &     37.57    &     41.87    &       0.9   \\
      \ion{Fe}{9} 189.941    &     19.13    &     19.09    &       1.0   \\
      \ion{Fe}{9} 197.862    &     23.97    &     21.97    &       1.1   \\
     \ion{Fe}{10} 184.536    &    256.62    &    198.67    &       1.3   \\
     \ion{Fe}{11} 180.401    &    948.90    &    892.12    &       1.1   \\
     \ion{Fe}{11} 188.216    &    422.43    &    424.65    &       1.0   \\
     \ion{Fe}{11} 188.299    &    291.92    &    261.12    &       1.1   \\
     \ion{Fe}{11} 192.813    &     90.06    &     89.19    &       1.0   \\
     \ion{Fe}{12} 192.394    &    141.05    &    150.80    &       0.9   \\
     \ion{Fe}{12} 193.509    &    323.34    &    318.25    &       1.0   \\
     \ion{Fe}{12} 195.119    &    367.57    &    470.87    &       0.8   \\
     \ion{Si}{10} 258.375    &    126.01    &    119.50    &       1.1   \\
     \ion{Si}{10} 261.058    &     68.07    &     76.63    &       0.9   \\ 
     \ion{Fe}{13} 202.044    &    236.04    &    186.15    &       1.3   \\
     \ion{Fe}{13} 203.826    &     44.11    &     37.34    &       1.2   \\
     \ion{Fe}{14} 211.316    &     37.02    &     35.94    &       1.0   \\
     \ion{Fe}{14} 270.519    &      8.04    &      8.93    &       0.9   \\
     \ion{Fe}{14} 274.203    &     21.33    &     20.36    &       1.0   \\
     \ion{Fe}{15} 284.160    &     17.20    &     17.25    &       1.0   \\
       \ion{O}{6} 184.117    &      9.01    &      8.83    &       1.0   
       \enddata
\tablenotetext{a}{Observed intensities are from 2007 November 4 and use the \citet{warren2014b}
  effective areas. The computed intensities are from the Gaussian DEM and $R$ is the ratio of
  observed to computed. The \ion{O}{6} line uses the scaled emissivity. The units for the
  intensities are erg cm$^2$ s$^{-1}$ sr$^{-1}$.}
  \label{table:offlimb}
\end{deluxetable}

\ion{O}{6} peaks at about $\log T = 5.6$ and emission from this ion is observed in the corona
because of the long tail on the ionization fraction.  In the transient brightenings and flares
that we observe we expect the emission from \ion{O}{6} to be formed much closer to the peak in the
ionization fraction. We now turn to observations in the quiet Sun, where this also expected to be
true.

The intensities of coronal emission lines in the on-disk, quiet Sun have been studied
by \citet{brooks2009}, who analyzed 45 full CCD observations taken during the 2007, solar minimum
between solar cycles 23 and 24. This analysis provided insights into the temperature structure of
the quiet corona and identified which emission lines and atomic data provided consistent
results. It did not, however, consider emission at transition region temperatures, which is very
weak in the quiet Sun.

We have taken several of the observations from the \citet{brooks2009} study and constructed
spatially and temporally averaged spectra. These spectra cover a region $90\arcsec\times90\arcsec$
in area with the 1\arcsec\ slit and 1\arcsec\ steps. Each exposure was 90\,s in duration. This
accumulates to about $7.3\times10^5$\,s of observing. Two of these quiet Sun spectra are shown in
Figure~\ref{fig:spec_disk}.

Inspection of the averaged spectra indicate that the relative line intensities are similar to what
is observed in the transient heating events. The high-FIP O lines are strong relative to the
low-FIP Si and Mg lines.

We have fit all of the relevant lines with Gaussians and computed calibrated line
intensities. The \ion{Fe}{13} and \ion{Mg}{7} ratios are used to compute electrons densities and
total pressures, which are much smaller than those observed in the active regions. The EM loci
curves are shown in Figure~\ref{fig:loci_disk} and are consistent with the simple inspection of
the spectra. The EM loci computed assuming photospheric abundances are more consistent than those
computed assuming a coronal composition. Even for photospheric abundances, however, there is an
offset between the O and Mg lines. The Mg lines are very weak and it difficult to assess the
uncertainties in the intensities.

The EM loci for \ion{O}{6} using both the computed emissivity from CHIANTI and the scaled
emissivity are shown in Figure~\ref{fig:loci_disk}. It is clear from these plots that without the
application of some sort of adjustment the \ion{O}{6} 184.117\,\AA\ line is not in agreement with
the other O lines or the Mg and Si lines. Scaling the emissivity for this line by a factor of 3.4
makes it more consistent with the EM loci curves for the other O lines. 
\appendix
\section{The EIS Calibration}

The relative calibration of EIS is an important component of this analysis. \citet{delzanna2013}
noted that some temperature and density insensitive line ratios that involved the long and short
wavelength detectors were not in agreement with theory. The \ion{Fe}{24} 192.04/255.10\,\AA\
ratio, for example, should be 2.5 but was observed to be above 4 and changing with
time. \citet{delzanna2013} and \cite{warren2014b} proposed modifications to the pre-flight
calibration that attempted to reconcile the available atomic data and the
observations. \cite{warren2014b} fit the time-dependent trends in the effective areas with
exponentials, which allows them to be extrapolated to the more recent observations that we have
analyzed here.

We have fit the \ion{Fe}{24} lines in each spatial pixel for both the 2015 June 21 and 2015 March
11 observations. Taking care to avoid spectra for which the 192.04\,\AA\ line is saturated, we
find median ratios of 2.61 and 2.51 using the revised calibration, which is in agreement with theory.
The pre-flight calibration yields ratios of 4.8 and 4.7.

A similar analysis of the \ion{Fe}{14} 274.203/211.316\,\AA\ branching ratio in these observations
yields less satisfactory results. The theoretical ratio from CHIANTI is 0.52. Using the pre-flight
calibration we obtain ratios of 0.19 and 0.18. Applying the revised calibration yields ratios of
0.78 and 0.79, which are about 50\% too high. Inspection of the EM loci curves for the recent
observations presented in Figures~\ref{fig:r20140316}--\ref{fig:r20150621} indicates generally
good agreement between the long and short wavelength lines. For example, \ion{Si}{7} 275.368
and \ion{Si}{10} 258.375\,\AA\ are generally in good agreement with the Fe lines formed at similar
temperatures. This suggests that the evolution of the EIS effective area near 211\,\AA\ is not
being properly accounted for in the revised calibration. Additional work is underway to evaluate
and correct this problem. 



\clearpage

\begin{thebibliography}{}
\expandafter\ifx\csname natexlab\endcsname\relax\def\natexlab#1{#1}\fi

\bibitem[{{Baker} {et~al.}(2015){Baker}, {Brooks}, {D{\'e}moulin}, {Yardley},
  {van Driel-Gesztelyi}, {Long}, \& {Green}}]{baker2015}
{Baker}, D., {Brooks}, D.~H., {D{\'e}moulin}, P., {et~al.} 2015, \apj, 802, 104

\bibitem[{{Brooks} {et~al.}(2015){Brooks}, {Ugarte-Urra}, \&
  {Warren}}]{brooks2015}
{Brooks}, D.~H., {Ugarte-Urra}, I., \& {Warren}, H.~P. 2015, Nature
  Communications, 6, 5947

\bibitem[{{Brooks} \& {Warren}(2011)}]{brooks2011}
{Brooks}, D.~H., \& {Warren}, H.~P. 2011, \apjl, 727, L13

\bibitem[{{Brooks} {et~al.}(2009){Brooks}, {Warren}, {Williams}, \&
  {Watanabe}}]{brooks2009}
{Brooks}, D.~H., {Warren}, H.~P., {Williams}, D.~R., \& {Watanabe}, T. 2009,
  \apj, 705, 1522

\bibitem[{{Caffau} {et~al.}(2011){Caffau}, {Ludwig}, {Steffen}, {Freytag}, \&
  {Bonifacio}}]{caffau2011}
{Caffau}, E., {Ludwig}, H.-G., {Steffen}, M., {Freytag}, B., \& {Bonifacio}, P.
  2011, \solphys, 268, 255

\bibitem[{{Culhane} {et~al.}(2007){Culhane}, {Harra}, {James}, {Al-Janabi},
  {Bradley}, {Chaudry}, {Rees}, {Tandy}, {Thomas}, {Whillock}, {Winter},
  {Doschek}, {Korendyke}, {Brown}, {Myers}, {Mariska}, {Seely}, {Lang}, {Kent},
  {Shaughnessy}, {Young}, {Simnett}, {Castelli}, {Mahmoud}, {Mapson-Menard},
  {Probyn}, {Thomas}, {Davila}, {Dere}, {Windt}, {Shea}, {Hagood}, {Moye},
  {Hara}, {Watanabe}, {Matsuzaki}, {Kosugi}, {Hansteen}, \&
  {Wikstol}}]{culhane2007}
{Culhane}, J.~L., {Harra}, L.~K., {James}, A.~M., {et~al.} 2007, \solphys, 243,
  19

\bibitem[{{De Pontieu} {et~al.}(2014){De Pontieu}, {Title}, {Lemen}, {Kushner},
  {Akin}, {Allard}, {Berger}, {Boerner}, {Cheung}, {Chou}, {Drake}, {Duncan},
  {Freeland}, {Heyman}, {Hoffman}, {Hurlburt}, {Lindgren}, {Mathur}, {Rehse},
  {Sabolish}, {Seguin}, {Schrijver}, {Tarbell}, {W{\"u}lser}, {Wolfson},
  {Yanari}, {Mudge}, {Nguyen-Phuc}, {Timmons}, {van Bezooijen}, {Weingrod},
  {Brookner}, {Butcher}, {Dougherty}, {Eder}, {Knagenhjelm}, {Larsen},
  {Mansir}, {Phan}, {Boyle}, {Cheimets}, {DeLuca}, {Golub}, {Gates}, {Hertz},
  {McKillop}, {Park}, {Perry}, {Podgorski}, {Reeves}, {Saar}, {Testa}, {Tian},
  {Weber}, {Dunn}, {Eccles}, {Jaeggli}, {Kankelborg}, {Mashburn}, {Pust},
  {Springer}, {Carvalho}, {Kleint}, {Marmie}, {Mazmanian}, {Pereira}, {Sawyer},
  {Strong}, {Worden}, {Carlsson}, {Hansteen}, {Leenaarts}, {Wiesmann},
  {Aloise}, {Chu}, {Bush}, {Scherrer}, {Brekke}, {Martinez-Sykora}, {Lites},
  {McIntosh}, {Uitenbroek}, {Okamoto}, {Gummin}, {Auker}, {Jerram}, {Pool}, \&
  {Waltham}}]{depontieu2014}
{De Pontieu}, B., {Title}, A.~M., {Lemen}, J.~R., {et~al.} 2014, \solphys, 289,
  2733

\bibitem[{{Del Zanna}(2013{\natexlab{a}})}]{delzanna2013b}
{Del Zanna}, G. 2013{\natexlab{a}}, \aap, 555, A47

\bibitem[{{Del Zanna}(2013{\natexlab{b}})}]{delzanna2013}
---. 2013{\natexlab{b}}, \aap, 558, A73

\bibitem[{{Del Zanna} {et~al.}(2015){Del Zanna}, {Dere}, {Young}, {Landi}, \&
  {Mason}}]{delzanna2015b}
{Del Zanna}, G., {Dere}, K.~P., {Young}, P.~R., {Landi}, E., \& {Mason}, H.~E.
  2015, \aap, 582, A56

\bibitem[{{Del Zanna} \& {Mason}(2014)}]{delzanna2014}
{Del Zanna}, G., \& {Mason}, H.~E. 2014, \aap, 565, A14

\bibitem[{{Dennis} {et~al.}(2015){Dennis}, {Phillips}, {Schwartz}, {Tolbert},
  {Starr}, \& {Nittler}}]{dennis2015}
{Dennis}, B.~R., {Phillips}, K.~J.~H., {Schwartz}, R.~A., {et~al.} 2015, \apj,
  803, 67

\bibitem[{{Dere} {et~al.}(1997){Dere}, {Landi}, {Mason}, {Monsignori Fossi}, \&
  {Young}}]{dere1997}
{Dere}, K.~P., {Landi}, E., {Mason}, H.~E., {Monsignori Fossi}, B.~C., \&
  {Young}, P.~R. 1997, \aaps, 125, 149

\bibitem[{{Doschek} {et~al.}(2015){Doschek}, {Warren}, \&
  {Feldman}}]{doschek2015}
{Doschek}, G.~A., {Warren}, H.~P., \& {Feldman}, U. 2015, \apjl, 808, L7

\bibitem[{{Dupree}(1972)}]{dupree1972}
{Dupree}, A.~K. 1972, \apj, 178, 527

\bibitem[{{Feldman}(1983)}]{feldman1983}
{Feldman}, U. 1983, \apj, 275, 367

\bibitem[{{Feldman} {et~al.}(1992){Feldman}, {Mandelbaum}, {Seely}, {Doschek},
  \& {Gursky}}]{feldman1992}
{Feldman}, U., {Mandelbaum}, P., {Seely}, J.~F., {Doschek}, G.~A., \& {Gursky},
  H. 1992, \apjs, 81, 387

\bibitem[{{Feldman} {et~al.}(1998){Feldman}, {Sch{\"u}hle}, {Widing}, \&
  {Laming}}]{feldman1998}
{Feldman}, U., {Sch{\"u}hle}, U., {Widing}, K.~G., \& {Laming}, J.~M. 1998,
  \apj, 505, 999

\bibitem[{{Fludra} \& {Schmelz}(1999)}]{fludra1999}
{Fludra}, A., \& {Schmelz}, J.~T. 1999, \aap, 348, 286

\bibitem[{{Freeland} \& {Handy}(1998)}]{freeland1998}
{Freeland}, S.~L., \& {Handy}, B.~N. 1998, \solphys, 182, 497

\bibitem[{{Hansteen} {et~al.}(2014){Hansteen}, {De Pontieu}, {Carlsson},
  {Lemen}, {Title}, {Boerner}, {Hurlburt}, {Tarbell}, {Wuelser}, {Pereira}, {De
  Luca}, {Golub}, {McKillop}, {Reeves}, {Saar}, {Testa}, {Tian}, {Kankelborg},
  {Jaeggli}, {Kleint}, \& {Mart{\'{\i}}nez-Sykora}}]{hansteen2014}
{Hansteen}, V., {De Pontieu}, B., {Carlsson}, M., {et~al.} 2014, Science, 346,
  arXiv:1412.3611

\bibitem[{{Kashyap} \& {Drake}(1998)}]{kashyap1998}
{Kashyap}, V., \& {Drake}, J.~J. 1998, \apj, 503, 450

\bibitem[{{Kashyap} \& {Drake}(2000)}]{kashyap2000}
---. 2000, Bulletin of the Astronomical Society of India, 28, 475

\bibitem[{{Kobayashi} {et~al.}(2011){Kobayashi}, {Cirtain}, {Golub},
  {Winebarger}, {Hertz}, {Cheimets}, {Caldwell}, {Korreck}, {Robinson},
  {Reardon}, {Kester}, {Griffith}, \& {Young}}]{kobayashi2011}
{Kobayashi}, K., {Cirtain}, J., {Golub}, L., {et~al.} 2011, in Society of
  Photo-Optical Instrumentation Engineers (SPIE) Conference Series, Vol. 8147,
  Society of Photo-Optical Instrumentation Engineers (SPIE) Conference Series,
  1

\bibitem[{{Korendyke} {et~al.}(2006){Korendyke}, {Brown}, {Thomas}, {Keyser},
  {Davila}, {Hagood}, {Hara}, {Heidemann}, {James}, {Lang}, {Mariska}, {Moser},
  {Moye}, {Myers}, {Probyn}, {Seely}, {Shea}, {Shepler}, \&
  {Tandy}}]{korendyke2006}
{Korendyke}, C.~M., {Brown}, C.~M., {Thomas}, R.~J., {et~al.} 2006, \ao, 45,
  8674

\bibitem[{{Laming}(2015)}]{laming2015}
{Laming}, J.~M. 2015, Living Reviews in Solar Physics, 12, 2

\bibitem[{{Laming} {et~al.}(1995){Laming}, {Drake}, \& {Widing}}]{laming1995}
{Laming}, J.~M., {Drake}, J.~J., \& {Widing}, K.~G. 1995, \apj, 443, 416

\bibitem[{{Landi} {et~al.}(2002){Landi}, {Feldman}, \& {Dere}}]{landi2002}
{Landi}, E., {Feldman}, U., \& {Dere}, K.~P. 2002, \apj, 574, 495

\bibitem[{{Landi} \& {Young}(2009)}]{landi2009}
{Landi}, E., \& {Young}, P.~R. 2009, \apj, 706, 1

\bibitem[{{Lang} {et~al.}(2006){Lang}, {Kent}, {Paustian}, {Brown}, {Keyser},
  {Anderson}, {Case}, {Chaudry}, {James}, {Korendyke}, {Pike}, {Probyn},
  {Rippington}, {Seely}, {Tandy}, \& {Whillock}}]{lang2006}
{Lang}, J., {Kent}, B.~J., {Paustian}, W., {et~al.} 2006, \ao, 45, 8689

\bibitem[{{Muglach} {et~al.}(2010){Muglach}, {Landi}, \&
  {Doschek}}]{muglach2010}
{Muglach}, K., {Landi}, E., \& {Doschek}, G.~A. 2010, \apj, 708, 550

\bibitem[{{Raymond} {et~al.}(1997){Raymond}, {Kohl}, {Noci}, {Antonucci},
  {Tondello}, {Huber}, {Gardner}, {Nicolosi}, {Fineschi}, {Romoli}, {Spadaro},
  {Siegmund}, {Benna}, {Ciaravella}, {Cranmer}, {Giordano}, {Karovska},
  {Martin}, {Michels}, {Modigliani}, {Naletto}, {Panasyuk}, {Pernechele},
  {Poletto}, {Smith}, {Suleiman}, \& {Strachan}}]{raymond1997}
{Raymond}, J.~C., {Kohl}, J.~L., {Noci}, G., {et~al.} 1997, \solphys, 175, 645

\bibitem[{{Sheeley}(1995)}]{sheeley1995}
{Sheeley}, Jr., N.~R. 1995, \apj, 440, 884

\bibitem[{{Sylwester} {et~al.}(2015){Sylwester}, {Phillips}, {Sylwester}, \&
  {K{\c e}pa}}]{sylwester2015}
{Sylwester}, B., {Phillips}, K.~J.~H., {Sylwester}, J., \& {K{\c e}pa}, A.
  2015, \apj, 805, 49

\bibitem[{{Teriaca} {et~al.}(2012){Teriaca}, {Andretta}, {Auch{\`e}re},
  {Brown}, {Buchlin}, {Cauzzi}, {Culhane}, {Curdt}, {Davila}, {Del Zanna},
  {Doschek}, {Fineschi}, {Fludra}, {Gallagher}, {Green}, {Harra}, {Imada},
  {Innes}, {Kliem}, {Korendyke}, {Mariska}, {Mart{\'{\i}}nez-Pillet},
  {Parenti}, {Patsourakos}, {Peter}, {Poletto}, {Rutten}, {Sch{\"u}hle},
  {Siemer}, {Shimizu}, {Socas-Navarro}, {Solanki}, {Spadaro}, {Trujillo-Bueno},
  {Tsuneta}, {Dominguez}, {Vial}, {Walsh}, {Warren}, {Wiegelmann}, {Winter}, \&
  {Young}}]{teriaca2012}
{Teriaca}, L., {Andretta}, V., {Auch{\`e}re}, F., {et~al.} 2012, Experimental
  Astronomy, 34, 273

\bibitem[{{Warren}(2005)}]{warren2005}
{Warren}, H.~P. 2005, \apjs, 157, 147

\bibitem[{{Warren}(2014)}]{warren2014a}
---. 2014, \apjl, 786, L2

\bibitem[{{Warren} \& {Brooks}(2009)}]{warren2009}
{Warren}, H.~P., \& {Brooks}, D.~H. 2009, \apj, 700, 762

\bibitem[{{Warren} {et~al.}(2014){Warren}, {Ugarte-Urra}, \&
  {Landi}}]{warren2014b}
{Warren}, H.~P., {Ugarte-Urra}, I., \& {Landi}, E. 2014, \apjs, 213, 11

\bibitem[{{Warren} {et~al.}(2010){Warren}, {Winebarger}, \&
  {Brooks}}]{warren2010}
{Warren}, H.~P., {Winebarger}, A.~R., \& {Brooks}, D.~H. 2010, \apj, 711, 228

\bibitem[{{Warren} {et~al.}(2012){Warren}, {Winebarger}, \&
  {Brooks}}]{warren2012}
---. 2012, \apj, 759, 141

\bibitem[{{Widing} \& {Feldman}(2001)}]{widing2001}
{Widing}, K.~G., \& {Feldman}, U. 2001, \apj, 555, 426

\bibitem[{{Young}(2005)}]{young2005}
{Young}, P.~R. 2005, \aap, 444, L45

\end{thebibliography}
\end{document}